\documentclass[fleqn,usenatbib]{mnras}

\usepackage[T1]{fontenc}
\usepackage{ae,aecompl}

\usepackage{graphicx}	
\usepackage{amsmath}	
\usepackage{amssymb}	

\newcommand{\NumBin}{4}
\newcommand{\NumFULL}{356}
\newcommand{\Numbf}{41}
\newcommand{\NumPRV}{20}
\newcommand{\NumNPRV}{38}
\newcommand{\NumNP}{238}
\newcommand{\Numu}{284}
\newcommand{\Numv}{306}
\newcommand{\Numw}{180}
\newcommand{\NumI}{118}
\newcommand{\NumP}{66}
\newcommand{\NumN}{52}
\newcommand{\NumG}{38}
\newcommand{\highsig}{15.5}
\newcommand{\lowsig}{1.0}
\newcommand{\medsig}{3.8}
\newcommand{\Nlace}{21}
\newcommand{\Nbaya}{20}

\newcommand{\G}{\textit{Gaia}}
\def\elem#1{\mathrm{#1}} 

\title[PARSEC III: Parallaxes of \NumI\ L \& T dwarfs]{Parallaxes of southern extremely cool
  objects\newline III: \NumI\ L \& T dwarfs}

\author[R. L. Smart et al.]{R.~L.    Smart$^{1}$\thanks{E-mail: smart@oato.inaf.it (RLS)},
B.       Bucciarelli$^{1}$,
H.~R.~A. Jones$^{2}$,
F.       Marocco$^{3}$,
A.~H.    Andrei$^{4}$,           \newauthor
B.       Goldman$^{5}$,
R.~A.    Mendez$^{6}$,
V.~A.  d'Avila$^{4}$,
B.       Burningham$^{2}$,
J.~I.~B. Camargo$^{4}$,            \newauthor
M.~T.    Crosta$^{1}$,
M.       Dapr\`{a}$^{7}$,
J.~S.    Jenkins$^{6}$,
R.       Lachaume$^{8, 5}$,
M.~G.    Lattanzi$^{1}$, \newauthor
J.~L.    Penna$^{4}$,          
D.~J.    Pinfield$^{2}$,
D.~N. da Silva Neto$^{4}$,
A.       Sozzetti$^{1}$        and 
A.       Vecchiato$^{1}$.
\\
$^{1}$Istituto Nazionale di Astrofisica, Osservatorio Astrofisico di Torino, Strada Osservatorio 20, 10025 Pino Torinese, Italy\\
$^{2}$Center for Astrophysics Research, University of Hertfordshire, Hatfield AL10 9AB,UK\\
$^{3}$Jet Propulsion Laboratory, California Institute of Technology, 4800 Oak Grove Dr., Pasadena, CA 91109, USA\\
$^{4}$Observat\'{o}rio Nacional/MCT,R, Gal. Jos\'{e} Cristino 77,CEP20921-400,RJ, Brazil\\
$^{5}$Max Planck Institute for Astronomy, Koenigstuhl 17,D--69117 Heidelberg,     Germany\\
$^{6}$Departamento de Astronomia, Universidad de Chile, Casilla 36-D, Santiago, Chile\\
$^{7}$Department of Physics and Astronomy, LaserLaB, Vrije Universiteit, De Boelelaan 1081, NL-1081 HV Amsterdam, Netherlands\\
$^{8}$Centro de Astroingenier\'ia, Instituto de Astrof\'isica, Pontificia Universidad Cat\'{o}lica de Chile, Vicu\~na Mackenna 4860, Macul, Santiago, Chile\\
}

\date{Accepted XXX. Received YYY; in original form ZZZ}

\pubyear{2018}

\begin{document}
\label{firstpage}
\pagerange{\pageref{firstpage}--\pageref{lastpage}}
\maketitle

\begin{abstract}
   {We present new results from the Parallaxes of Southern Extremely Cool
     dwarfs program to measure parallaxes, proper motions and multi-epoch
     photometry of L and early T dwarfs.  The observations were made on 108
     nights over the course of 8 years using the Wide Field Imager on the ESO
     2.2m telescope. We present \NumI\ new parallaxes of L \& T dwarfs of
     which \NumN\ have no published values and 24 of the \NumP\ published
     values are preliminary estimates from this program. The parallax
     precision varies from \lowsig\ to \highsig\,mas with a median of
     \medsig\,mas. We find evidence for 2 objects with long term photometric
     variation and 24 new moving group candidates. 
     We cross-match our sample to published photometric
     catalogues and find standard magnitudes in up to 16 pass-bands from which
     we build spectral energy distributions and H-R diagrams. This allows us
     to confirm the theoretically anticipated minimum in radius between stars
     and brown dwarfs across the hydrogen burning minimum mass. We find the
     minimum occurs between L2 and L6 and verify the predicted steep
     dependence of radius in the hydrogen burning regime and the gentle rise
     into the degenerate brown dwarf regime. We find a relatively young
     age of $\sim$2~\,Gyr from the kinematics of our sample.}
\end{abstract}




\section{Introduction}

Objects with spectral types L and T cover the mass range from the lowest mass
hydrogen burning stars, through slowly cooling sub-stellar objects down to
massive Jupiter type objects. Since the first tentative discoveries 30 years
ago \citep{1988Natur.336..656B,1989Natur.339...38L} over 3000 are known today
and this number will increase exponentially with the planned deep optical and
infrared surveys (e.g. with the Large Synoptic Survey Telescope --
\citealt{2017arXiv170804058L}; the Panoramic Survey Telescope and Rapid 
Response System -- \citealt{2016arXiv161205560C}; the Wide Field Infrared Survey
Telescope -- \citealt{2015arXiv150303757S}; and Euclid --
\citealt{2010SPIE.7731E..1HL}). Observations and statistical studies of these
objects can be used to constrain proposed stellar/sub-stellar formation
processes, local galactic kinematics, understanding giant planet atmospheres
and mapping the stellar to sub-stellar boundary. The lower mass sub-stellar
objects are continually cooling and therefore changing with time which,
combined with their ubiquity, make them promising galactic chronometers. To
realize their promise a large sample with measured distances is needed to
enable a complete calibration.

\begin{figure}
\begin{center}
\includegraphics[scale=0.45]{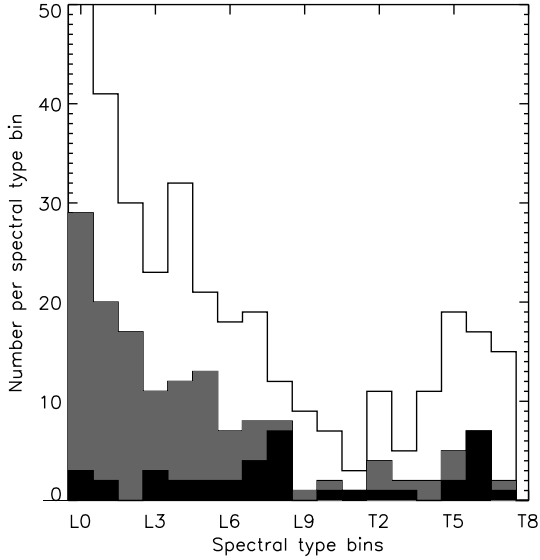}
\caption{Distribution of parallaxes of L0 to T8 dwarfs with spectral type.
  The black area represents the \Numbf\ objects published before 2007, the
  grey area the \NumI\ PARSEC objects and the white area represents all
  objects published today, a total of \NumFULL\ L0 to T8 objects.  There is an
  overlap of \NumP\ objects between the PARSEC and published objects as
  discussed in Section~\ref{ComparisontoPublishedParallaxes}.}
\label{histrogram}
\end{center}
\end{figure}

The Parallaxes of Southern Extremely Cool dwarfs (hereafter PARSEC) program
was instigated to generate a large sample of these objects with measured
parallaxes. In 2007 only \Numbf\ L0 to T8 objects had published parallaxes,
and in PARSEC we aimed to increase the sample of objects to at least 10 for
each L dwarf spectral sub-type, and, included bright southern T dwarfs to
increment the T dwarf coverage. In this contribution we report the PARSEC
parallaxes of \NumI\ L0 to T8 dwarfs which, combined with objects with
literature parallaxes (described in Section~\ref{Astrometricresults}), brings
the total to \NumFULL\ objects distributed in spectral type as shown in
Fig.~\ref{histrogram} which we refer to as the Full Sample.

For each L dwarf sub-type the number of objects is now at least 10, except L9 where
we have 9. As discussed in
Section~\ref{conclusion} the ESA \G\ mission, that is measuring parallaxes for
$10^9$ objects, will provide a significant numbers of early L dwarfs but will
have only a few late L dwarfs and less that 10 T dwarfs, so the cooler objects
will remain the domain of small field pointed programs.  In 2010 a complementary
program to PARSEC was started on the ESO New Technology Telescope targeting
late T dwarfs \citep{2013MNRAS.433.2054S} that are too faint for PARSEC or \G.

Preliminary results for the PARSEC program have been published in
\cite{2011AJ....141...54A} and \cite{2013AJ....146..161M} using observations
from the first 2-3 years; here we provide results from the full program with
observations covering 8 years. In Section~2 we briefly present the PARSEC
program, in Section~3 we present the astrometric results and in Section 4 and
5 we present applications of these results combined with literature measures
to the problem of absolute magnitude calibration, local kinematics and the
location of the stellar - brown dwarf boundary.

\section{The PARSEC program}

The instrument, observational procedures, reduction procedures and target
selection is described in detail in  \cite{2011AJ....141...54A}, here we
briefly summarize the main points.

\subsection{Telescope and detector}
The PARSEC observations where made on the Wide Field Imager
\citep[WFI,][]{1999Msngr..95...15B} of the ESO MPIA 2.2m telescope. This is a
mosaic of 8 EEV CCD44 chips with 2k$\times$4k 15\,$\mu$m pixels, for the
results presented here we only used observations from the top half of CCD\#7
(Priscilla). Limiting our reductions to this region was a balance between
simplicity of the required astrometric transforms - the larger the adopted
area being modeled the more complicated transforms were required - and number
of anonymous reference stars for which we required a minimum of 12. 

This telescope and instrument combination has a number of positive characteristics:
\begin{enumerate}
\item The camera is fixed and stable, crucial for small field relative
  astrometry.
\item The 0.2\,\arcsec/pixel focal plane scale allows at least 2 pixels per
  full width half maximum (e.g.  Nyquist sampling) even in the best seeing.
\item The total field size of 0.3\,sq.\,deg. provides a large field to search
  for nearby companions.
\item This combination had already been used for the determination of
  parallaxes \citep{2007AA...470..387D}.
\end{enumerate}
We observed all objects in the $z$ band (Z+$/$61 ESO\#846, central wavelength
964.8\,$\mu$m, FWHM 61.6\,$\mu$m) which provided the best ratio of exposure
time and signal-to-noise for these very red targets.

\subsection{Observational procedure}

Each observation consisted of a short exposure to visually locate the target
and then an application of the WFI move-to-pixel procedure to move the target
to pixel 3400,3500 in CCD\#7. We then made two exposures of 150\,s for objects
with $z\le$18.0 and 300\,s for $z\ge$18.0 offset by 24\,pixels in both
axes. If we found the signal-to-noise of the first exposure to be less than
100 we increased the exposure time of the second accordingly. The short
location frames were saved and used in the reduction process to model the $z$
band fringing.

\begin{table*}
 \caption{Targets, published magnitudes, spectral types, radial velocities and parallaxes}
 \label{baseinfo}
 \begin{tabular}{llllllll}    
  \hline
Target    & Discovery Name$^{\rm Ref.}$      & $z$ band  & Optical.    & NIR        &  RV, $\sigma_{RV}$$^{\rm Ref.}$ & $\varpi, \sigma_{\varpi}$$^{\rm Ref.}$& Multiple \\
Code      &                             & ~~~~mag   & SpT$^{\rm Ref.}$ & SpT$^{\rm Ref.}$ &      km/s                 & ~~~~~ mas                & Code$^{\rm Ref.}$ \\ 
\hline                                                   
0004-4044  &          GJ1001B$^{2}$ &    15.8 & L5$^{14}$ & L4.5$^{39}$ &      32.9,      0.2$^{71}$ &      77.0,      2.1$^{83}$ &   VB;$^{40}$\\
0013-2235  & 2MASSIJ0013578-223520$^{27}$ &    18.6 & L4$^{27}$ & L5.5$^{85}$ &       ...                &       ...                &$^{}$\\
0016-4056  & 2MASS00165953-4056541$^{63}$ &    18.0 & L3.5$^{63}$ & ...$^{}$ &       ...                &       ...                &$^{}$\\
0032-4405  & EROS-MPJ0032-4405$^{2}$ &    17.1 & L0$\gamma$$^{65}$ & L0$\gamma$$^{77}$ &       ...                &      38.4,      4.8$^{74}$ &   MG;$^{84}$\\
0034-0706  & 2MASSIJ0034568-070601$^{27}$ &    18.2 & L3$^{27}$ & L4.5$^{85}$ &       ...                &       ...                &$^{}$\\
0054-0031  & SDSSpJ005406.55-003101.8$^{18}$ &    18.3 & L1$^{19}$ & L2$^{85}$ &      -5.7,     13.0$^{69}$ &       ...                &$^{}$\\
0058-0651  & 2MASSWJ0058425-065123$^{9}$ &    17.1 & L0$^{9}$ & L0$^{85}$ &       ...                &      33.8,      4.0$^{76}$ &   MG;$^{84}$\\
0109-5100  & 2MASS01090150-5100494$^{17}$ &    14.6 & M8.5$^{44}$ & L2$^{44}$ &       ...                &      57.8,      3.3$^{76}$ &$^{}$\\
0117-3403  & 2MASSIJ0117474-340325$^{31}$ &    17.9 & L2:$^{31}$ & L1$\beta$$^{77}$ &       ...                &      26.1,      1.9$^{}$ &   MG;$^{84}$\\
0128-5545  & 2MASS01282664-5545343$^{54}$ &    16.6 & L2$^{60}$ & L1$^{54}$ &       ...                &       ...                &$^{}$\\
0144-0716  & 2MASS01443536-0716142$^{28}$ &    16.9 & L5$^{28}$ & L5$^{76}$ &      -2.6,      0.1$^{71}$ &       ...                &$^{}$\\
0147-4954  & 2MASSJ01473282-4954478$^{49}$ &    15.8 & ...$^{}$ & L2.0$^{76}$ &       ...                &       ...                &$^{}$\\
0205-1159  & DENIS-PJ0205.4-1159$^{1}$ &    17.4 & L7$^{5}$ & L5.5$^{39}$ &       ...                &      54.3,      1.6$^{88}$ &   UR;$^{45}$\\
0219-1938  &  SSSPMJ0219-1939$^{17}$ &    16.9 & L1$^{44}$ & L2.5$^{44}$ &       ...                &      37.2,      4.1$^{76}$ &$^{}$\\
0227-1624  & 2MASS02271036-1624479$^{60}$ &    16.1 & L1$^{60}$ & L0.5:$^{76}$ &      48.5,      0.2$^{71}$ &       ...                &$^{}$\\
0230-0953  & DENISJ02304500-0953050$^{68}$ &    17.7 & L0$^{68}$ & L1$^{85}$ &       ...                &      32.4,      3.7$^{76}$ &$^{}$\\
0235-0849  & 2MASS02354756-0849198$^{19}$ &    18.3 & L2$^{19}$ & L2:$^{85}$ &      22.8,      6.1$^{69}$ &       ...                &$^{}$\\
0235-2331  &          GJ1048B$^{13}$ &    15.2 & L1$^{13}$ & L1$^{13}$ &      15.4,      0.1$^{71}$ &      47.0,      0.9$^{55}$ &   VB;$^{13}$\\
0239-1735  & 2MASSIJ0239424-173547$^{31}$ &    16.6 & L0$^{31}$ & M9$^{76}$ &       ...                &      32.1,      4.7$^{76}$ &$^{}$\\
0255-4700  & DENIS-PJ0255-4700$^{4}$ &    16.1 & L8$^{63}$ & L9$^{51}$ &       ...                &     205.8,      0.5$^{88}$ &$^{}$\\
0257-3105  & 2MASS02572581-3105523$^{63}$ &    17.6 & L8$^{63}$ & L8:$^{76}$ &       ...                &      99.7,      6.7$^{76}$ &$^{}$\\
0318-3421  & 2MASS03185403-3421292$^{63}$ &    18.5 & L7$^{63}$ & ...$^{}$ &       ...                &      72.9,      7.7$^{74}$ &$^{}$\\
0357-0641  & 2MASS03572110-0641260$^{19}$ &    18.3 & L0$^{19}$ & ...$^{}$ &      89.7,     41.1$^{59;69}$ &       ...                &$^{}$\\
0357-4417  & DENIS-PJ035726.9-441730$^{30}$ &    16.7 & L0$\beta$$^{63}$ & L2p$^{76}$ &       ...                &       ...                &   MG;$^{84}$\\
0408-1450  & 2MASSIJ0408290-145033$^{35}$ &    16.9 & L2$^{31}$ & L4.5$^{35}$ &       ...                &       ...                &$^{}$\\
0423-0414  & SDSSpJ042348.57-041403.5$^{22}$ &    17.3 & L7.5$^{31}$ & T0$^{51}$ &       ...                &      67.5,      2.3$^{88}$ &   UR;$^{51}$\\
0439-2353  & 2MASSIJ0439010-235308$^{31}$ &    17.3 & L6.5$^{31}$ & L4.5$^{82}$ &       ...                &     110.4,      4.0$^{74}$ &$^{}$\\
0518-2828  & 2MASS05185995-2828372$^{41}$ &    18.8 & L7$^{63}$ & L6+T4$^{53}$ &       ...                &      43.7,      0.8$^{75}$ &   UR;$^{41;53}$\\
0523-1403  & 2MASSIJ0523382-140302$^{31}$ &    15.9 & L2.5$^{31}$ & L5$^{35}$ &      12.2,      0.1$^{71}$ &      80.9,      1.8$^{83}$ &$^{}$\\
0539-0059  & SDSSpJ053951.99-005902.0$^{8}$ &    16.7 & L5$^{8}$ & L5$^{39}$ &      13.9,      0.2$^{71}$ &      79.2,      1.0$^{88}$ &$^{}$\\
0559-1404  & 2MASS05591914-1404488$^{10}$ &    17.3 & T5$^{32}$ & T4.5$^{51}$ &       ...                &      96.8,      1.2$^{88}$ &$^{}$\\
0614-2019  &   SIPSJ0614-2019$^{73}$ &    17.6 & ...$^{}$ & L4$^{73}$ &       ...                &      34.3,      3.0$^{76}$ &$^{}$\\
0624-4521  & 2MASS06244595-4521548$^{60}$ &    17.2 & L5:$^{60}$ & L5$^{85}$ &       ...                &      81.2,      0.4$^{88}$ &$^{}$\\
0639-7418  & 2MASS06395596-7418446$^{56}$ &    18.5 & L5$^{56}$ & ...$^{}$ &       ...                &       ...                &$^{}$\\
0641-4322  & 2MASS06411840-4322329$^{60}$ &    16.3 & L1.5$^{60}$ & L2.5:$^{85}$ &       ...                &      51.1,      0.5$^{88}$ &$^{}$\\
0719-5051  & 2MASS07193188-5051410$^{60}$ &    16.5 & L0$^{60}$ & L0:$^{76}$ &       ...                &      34.6,      2.2$^{76}$ &$^{}$\\
0729-7843  & 2MASSJ07291084-7843358$^{73}$ &    18.3 & ...$^{}$ & L0.0$^{76}$ &       ...                &       ...                &$^{}$\\
0828-1309  &  SSSPMJ0829-1309$^{24}$ &    15.6 & L2$^{44}$ & L2:$^{76}$ &      25.8,      0.1$^{71}$ &      85.8,      0.1$^{81}$ &$^{}$\\
0832-0128  & 2MASSWJ0832045-012835$^{9}$ &    16.6 & L1.5$^{9}$ & L1$^{85}$ &      20.0,      1.3$^{26}$ &      40.4,      1.8$^{88}$ &$^{}$\\
0835-0819  & 2MASSIJ0835425-081923$^{31}$ &    15.9 & L5$^{31}$ & L4$^{76}$ &       ...                &     137.5,      0.4$^{88}$ &$^{}$\\
0859-1949  & 2MASSIJ0859254-194926$^{31}$ &    18.4 & L6::$^{31}$ & L8$^{80}$ &       ...                &      65.4,      6.1$^{74}$ &$^{}$\\
0909-0658  & DENIS-PJ0909-0658$^{3}$ &    16.2 & L0$^{63}$ & L0$^{85}$ &       ...                &      42.5,      4.2$^{73}$ &$^{}$\\
0921-2104  & 2MASS09211410-2104446$^{60}$ &    15.5 & L1.5$^{60}$ & L4p(blue)$^{61}$ &       ...                &       ...                &$^{}$\\
0922-8010  & 2MASS09221952-8010399$^{60}$ &    18.1 & L2::$^{60}$ & ...$^{}$ &       ...                &       ...                &$^{}$\\
0928-1603  & 2MASSWJ0928397-160312$^{9}$ &    18.1 & L2$^{9}$ & L2:$^{76}$ &       ...                &      34.4,      3.9$^{76}$ &$^{}$\\
1004-1318  & DENISJ1004403-131818$^{68}$ &    17.6 & L0$^{68}$ & L1:$^{76}$ &       ...                &       ...                &$^{}$\\
1004-3335  & 2MASSWJ1004392-333518$^{23}$ &    17.3 & L4$^{23}$ & L4.5:$^{85}$ &       ...                &      54.8,      5.6$^{73}$ &   VB;$^{47}$\\
1018-2909  & 2MASSWJ1018588-290953$^{23}$ &    16.7 & L1$^{23}$ & L0.5$^{85}$ &       ...                &      35.3,      3.2$^{73}$ &$^{}$\\
1045-0149  & 2MASSIJ1045240-014957$^{23}$ &    15.7 & L1$^{23}$ & L2$^{76}$ &       ...                &      61.8,      1.5$^{}$ &   MG;$^{64}$\\
1047-1815  & DENIS-PJ1047-1815$^{4}$ &    17.0 & L2.5$^{4}$ & L0.5$^{85}$ &       ...                &      37.9,      1.9$^{}$ &$^{}$\\
1058-1548  & DENIS-PJ1058.7-1548$^{1}$ &    16.9 & L3$^{5}$ & L3$^{39}$ &       ...                &      55.9,      0.6$^{88}$ &   MG;$^{86}$\\
1059-2113  & 2MASSIJ1059513-211308$^{31}$ &    17.1 & L1$^{31}$ & L3:$^{85}$ &       ...                &       ...                &$^{}$\\
1122-3512  & 2MASS11220826-3512363$^{46}$ &    18.1 & ...$^{}$ & T2$^{51}$ &       ...                &       ...                &$^{}$\\
1122-3916  & 2MASSWJ1122362-391605$^{23}$ &    18.4 & L3$^{23}$ & L3.5::$^{85}$ &       ...                &       ...                &$^{}$\\
1126-5003  & 2MASS11263991-5003550$^{58}$ &    15.9 & L4.5$^{61}$ & L6.5p$^{61}$ &       ...                &      60.8,      2.0$^{88}$ &$^{}$\\
1154-3400  & 2MASS11544223-3400390$^{30}$ &    16.6 & L0$^{63}$ & L0.5$^{85}$ &       ...                &       ...                &   MG;$^{87}$\\
1225-2739  & 2MASS12255432-2739466$^{6}$ &    18.8 & T6$^{32}$ & T6$^{51}$ &       ...                &      75.1,      2.5$^{29}$ &   UR;$^{33}$\\
1228-1547  & DENIS-PJ1228.2-1547$^{1}$ &    17.2 & L5$^{5}$ & L6$^{39}$ &      19.4,      5.0$^{72}$ &      44.8,      1.8$^{75}$ &   UR;$^{30}$\\
1246-3139  & WISEJ124629.65-313934.2$^{73}$ &    18.2 & ...$^{}$ & T2:$^{78}$ &       ...                &      87.3,      3.2$^{76}$ &$^{}$\\
1254-0122  & SDSSpJ125453.90-012247.4$^{12}$ &    18.0 & T2$^{32}$ & T2$^{51}$ &       ...                &      84.9,      1.9$^{20}$ &$^{}$\\
1326-2729  & 2MASSWJ1326201-272937$^{23}$ &    18.6 & L5$^{23}$ & L6.5:$^{85}$ &       ...                &       ...                &   MG;$^{87}$\\
\hline                                                 
\end{tabular}
\end{table*}

\begin{table*}
 \contcaption{Targets, published magnitudes, spectral types, radial velocities and parallaxes}
 \begin{tabular}{llllllll}    
  \hline
Target    & Discovery Name$^{\rm Ref.}$      & $z$ band  & Optical.    & NIR        &  RV, $\sigma_{RV}$$^{\rm Ref.}$ & $\pi, \sigma_{\pi}$$^{\rm Ref.}$& Multiple \\
Code      &                             & ~~~~mag   & SpT$^{\rm Ref.}$ & SpT$^{\rm Ref.}$ &      km/s                 & ~~~~~ mas                & Code$^{\rm Ref.}$ \\ 
\hline
1331-0116  & 2MASS13314894-0116500$^{19}$ &    18.4 & L6$^{19}$ & L8p(blue)$^{39}$ &       ...                &      67.3,     12.6$^{76}$ &$^{}$\\
1341-3052  & 2MASS13411160-3052505$^{60}$ &    17.3 & L2::$^{60}$ & L2.5:$^{85}$ &      33.7,      5.0$^{72}$ &       ...                &   SB;$^{85}$\\
1404-3159  & 2MASS14044948-3159330$^{57}$ &    18.8 & T0$^{62}$ & T2.5$^{57}$ &       ...                &      42.1,      1.1$^{75}$ &   UR;$^{62}$\\
1425-3650  & DENIS-PJ142527.97-365023.4$^{36}$ &    16.5 & L3:$^{60}$ & L5$^{36}$ &       5.3,      0.3$^{71}$ &      86.4,      0.8$^{83}$ &   MG;$^{86}$\\
1438-1309  & 2MASSWJ1438549-130910$^{9}$ &    18.2 & L3:$^{9}$ & L3$^{85}$ &       ...                &       ...                &$^{}$\\
1441-0945  &        G124-62BC$^{4}$ &    16.4 & L0.5$^{63}$ & L0.5$^{85}$ &       ...                &      30.7,      0.7$^{88}$ &   VB;$^{30;43}$\\
1457-2121  &       Gliese570D$^{11}$ &    18.8 & T7$^{32}$ & T7.5$^{51}$ &       ...                &     167.6,      4.6$^{88}$ &   VB;$^{11}$\\
1507-1627  & 2MASSWJ1507476-162738$^{7}$ &    15.6 & L5$^{9}$ & L5.5$^{39}$ &     -39.8,      0.1$^{71}$ &     133.9,      0.6$^{88}$ &$^{}$\\
1520-4422B &  WDSJ15200-4423B$^{54}$ &    16.0 & ...$^{}$ & L4.5$^{51}$ &       ...                &       ...                &$^{}$\\
1523-2347  & 2MASS15230657-2347526$^{54}$ &    17.0 & ...$^{}$ & L2.5$^{54}$ &       ...                &       ...                &$^{}$\\
1530-8145  & 2MASSJ15302867-8145375$^{37}$ &    17.0 & ...$^{}$ & L0.0$^{76}$ &       ...                &       ...                &$^{}$\\
1534-2952  & 2MASSIJ1534498-295227$^{21}$ &    18.4 & T6$^{32}$ & T5.5$^{51}$ &       ...                &      62.4,      1.3$^{75}$ &   UR;$^{34}$\\
1539-0520  & DENIS-PJ153941.96-052042.4$^{36}$ &    16.6 & L4:$^{63}$ & L2$^{36}$ &       ...                &      60.1,      1.2$^{88}$ &$^{}$\\
1547-2423  & 2MASS15474719-2423493$^{60}$ &    16.3 & M9p$^{60}$ & L0Int-G$^{77}$ &       ...                &      30.0,      1.1$^{}$ &   MG;$^{87}$\\
1548-1636  & 2MASS15485834-1636018$^{54}$ &    16.7 & ...$^{}$ & L2:$^{54}$ &       ...                &       ...                &$^{}$\\
1618-1321  & 2MASS16184503-1321297$^{63}$ &    16.6 & L0:$^{63}$ & M9.5$^{85}$ &       ...                &      21.9,      1.3$^{88}$ &$^{}$\\
1620-0416  &         GJ618.1B$^{15}$ &    18.0 & L2.5$^{15}$ & L2.5$^{85}$ &       ...                &      29.9,      2.7$^{55}$ &   VB;$^{15}$\\
1633-0640  & 2MASS16335933-0640552$^{48}$ &    19.0 & ...$^{}$ & L6$^{48}$ &       ...                &       ...                &$^{}$\\
1636-0034  & SDSSpJ163600.79-003452.6$^{8}$ &    17.0 & L0$^{8}$ & M9$^{85}$ &      -7.4,      4.1$^{59;69}$ &       ...                &$^{}$\\
1645-1319  & 2MASSWJ1645221-131951$^{23}$ &    15.0 & L1.5$^{23}$ & ...$^{}$ &       ...                &      89.3,      0.4$^{88}$ &$^{}$\\
1705-0516  & DENIS-PJ170548.38-051645.7$^{36}$ &    16.1 & ...$^{}$ & L4$^{36}$ &       ...                &      53.5,      1.0$^{}$ &   UR;$^{49}$\\
1707-0558  & 2MASS17072343-0558249$^{50}$ &    16.7 & ...$^{}$ & L3$^{50}$ &       ...                &       ...                &   UR;MG;$^{50}$\\
1750-0016  & 2MASS17502484-0016151$^{54}$ &    16.0 & ...$^{}$ & L5.5$^{54}$ &       ...                &     108.8,      0.8$^{88}$ &$^{}$\\
1753-6559  & 2MASS17534518-6559559$^{60}$ &    16.9 & L4::$^{60}$ & L4:$^{76}$ &       ...                &      58.0,      4.9$^{76}$ &$^{}$\\
1828-4849  & 2MASS18283572-4849046$^{42}$ &    18.7 & ...$^{}$ & T5.5$^{51}$ &       ...                &      87.9,      2.0$^{79}$ &$^{}$\\
1840-5631  & 2MASSJ18401904-5631138$^{73}$ &    18.9 & ...$^{}$ & L9.0$^{76}$ &       ...                &       ...                &$^{}$\\
1928-4356  & 2MASS19285196-4356256$^{60}$ &    17.9 & L4$^{60}$ & L4p$^{76}$ &       ...                &       ...                &$^{}$\\
1936-5502  & 2MASS19360187-5502322$^{60}$ &    17.2 & L5:$^{60}$ & L4$^{76}$ &       ...                &      43.3,      4.5$^{76}$ &$^{}$\\
1956-1754  & 2MASS19561542-1754252$^{54}$ &    16.1 & M8$^{60}$ & L0:$^{54}$ &       ...                &       ...                &$^{}$\\
2002-0521  & 2MASS20025073-0521524$^{56}$ &    18.2 & L6$^{56}$ & L7::$^{85}$ &       ...                &       ...                &$^{}$\\
2011-6201  & 2MASSJ20115649-6201127$^{73}$ &    18.8 & ...$^{}$ & sdM8$^{76}$ &       ...                &       ...                &$^{}$\\
2023-5946  & 2MASSJ20232858-5946519$^{73}$ &    18.7 & ...$^{}$ & M8.0$^{76}$ &       ...                &       ...                &$^{}$\\
2026-2943  & 2MASS20261584-2943124$^{56}$ &    17.3 & L1:$^{56}$ & L1+T6:$^{70}$ &       ...                &       ...                &   UR;$^{70;85}$\\
2041-3506  & 2MASS20414283-3506442$^{56}$ &    17.6 & L2:$^{56}$ & L2$^{85}$ &       ...                &       ...                &   MG;$^{86}$\\
2045-6332  &    SIPS2045-6332$^{76}$ &    15.4 & ...$^{}$ & L1:$^{76}$ &       ...                &      41.7,      1.5$^{83}$ &   MG;$^{72}$\\
2057-0252  &       2MUCD12054$^{31}$ &    15.6 & L1.5$^{31}$ & L1.5$^{36}$ &     -24.6,      0.4$^{71}$ &      64.7,      0.8$^{88}$ &$^{}$\\
2101-2944  & 2MASS21015233-2944050$^{76}$ &    18.8 & ...$^{}$ & L1$^{76}$ &       ...                &       ...                &$^{}$\\
2104-1037  & 2MASSIJ2104149-103736$^{31}$ &    16.6 & L2.5$^{63}$ & ...$^{}$ &       ...                &      57.2,      0.9$^{88}$ &$^{}$\\
2107-4544  & 2MASS21075409-4544064$^{60}$ &    17.3 & L0:$^{60}$ & L2.5$^{85}$ &       ...                &       ...                &$^{}$\\
2130-0845  & 2MASSWJ2130446-084520$^{63}$ &    16.7 & L1.5$^{63}$ & M8.5$^{85}$ &       ...                &       ...                &$^{}$\\
2132-1452  & 2MASS21324898-1452544$^{76}$ &    19.0 & ...$^{}$ & T4$^{76}$ &       ...                &       ...                &$^{}$\\
2150-7520  & 2MASS21501592-7520367$^{60}$ &    16.6 & L1:$^{60}$ & ...$^{}$ &       ...                &       ...                &$^{}$\\
2157-5534  & 2MASS21574904-5534420$^{60}$ &    17.0 & L0::$^{60}$ & ...$^{}$ &       ...                &       ...                &$^{}$\\
2158-1550  & 2MASS21580457-1550098$^{63}$ &    17.8 & L4:$^{63}$ & L4.5:$^{85}$ &       ...                &       ...                &$^{}$\\
2204-5646  &       epsIndiBab$^{25}$ &    16.7 & ...$^{}$ & T1+T6$^{51}$ &       ...                &     275.3,      3.0$^{88}$ &   VB;$^{16}$\\
2206-4217  & 2MASSWJ2206450-421721$^{9}$ &    18.3 & L2$^{9}$ & L4::$^{85}$ &       ...                &       ...                &$^{}$\\
2209-2711  & 2MASS22092183-2711329$^{76}$ &    18.9 & ...$^{}$ & T2.5$^{76}$ &       ...                &      47.9,     12.5$^{76}$ &$^{}$\\
2213-2136  & 2MASS22134491-2136079$^{56}$ &    17.9 & L0Int-G$^{63}$ & L0Int-G$^{77}$ &       ...                &      20.9,      1.9$^{}$ &$^{}$\\
2224-0158  & 2MASSWJ2224438-015852$^{9}$ &    16.9 & L4.5$^{9}$ & L3.5$^{39}$ &     -37.6,      0.1$^{71}$ &      86.1,      0.9$^{88}$ &$^{}$\\
2252-1730  & DENIS-PJ225210.73-173013.4$^{36}$ &    17.2 & ...$^{}$ & L7.5$^{36}$ &       ...                &      63.2,      1.6$^{75}$ &   UR;$^{52}$\\
2254-2840  & 2MASSIJ2254519-284025$^{31}$ &    16.5 & L0.5$^{31}$ & L0.5$^{36}$ &       ...                &       ...                &$^{}$\\
2255-0034  & SDSSpJ225529.09-003433.4$^{18}$ &    18.0 & L0:$^{18}$ & M8.5$^{85}$ &      12.3,     24.0$^{59;69}$ &      16.2,      2.6$^{38}$ &$^{}$\\
2310-1759  &  SSSPMJ2310-1759$^{17}$ &    16.9 & L0:$^{56}$ & L1$^{44}$ &       ...                &      36.4,      6.9$^{76}$ &$^{}$\\
2318-1301  & 2MASS23185497-1301106$^{76}$ &    18.8 & ...$^{}$ & T5$^{76}$ &       ...                &       ...                &$^{}$\\
2330-0347  & 2MASS23302258-0347189$^{56}$ &    17.0 & L1:$^{56}$ & L0.5$^{85}$ &       ...                &       ...                &$^{}$\\
2346-5928  &    SIPS2346-5928$^{73}$ &    17.3 & ...$^{}$ & L5.0$^{76}$ &       ...                &       ...                &$^{}$\\
2351-2537  & 2MASS23515044-2537367$^{67}$ &    14.8 & L0.5$^{67}$ & ...$^{}$ &     -10.0,      3.0$^{66}$ &       ...                &$^{}$\\
\hline                                                 
\end{tabular}
\end{table*}

\begin{table*}
 \contcaption{Targets, published magnitudes, spectral types, radial velocities and parallaxes}
 \begin{tabular}{llllllll}    
  \hline
\end{tabular}
\raggedright{Multiple Code: VB=Visual Binary, UR=Unresolved Binary, MG=Moving group, Bi=Binary, SB=Spectral Binary \newline References 1:\cite{1997A&A...327L..25D}, 2:\cite{1999A&A...351L...5G}, 3:\cite{1999A&AS..135...41D}, 4:\cite{1999AJ....118.2466M}, 5:\cite{1999ApJ...519..802K}, 6:\cite{1999ApJ...522L..65B}, 7:\cite{2000AJ....119..369R}, 8:\cite{2000AJ....119..928F}, 9:\cite{2000AJ....120..447K}, 10:\cite{2000AJ....120.1100B}, 11:\cite{2000ApJ...531L..57B}, 12:\cite{2000ApJ...536L..35L}, 13:\cite{2001AJ....121.2185G}, 14:\cite{2001AJ....121.3235K}, 15:\cite{2001AJ....122.1989W}, 16:\cite{2001AJ....122.3466M}, 17:\cite{2002A&A...389L..20L}, 18:\cite{2002AJ....123..458S}, 19:\cite{2002AJ....123.3409H}, 20:\cite{2002AJ....124.1170D}, 21:\cite{2002ApJ...564..421B}, 22:\cite{2002ApJ...564..466G}, 23:\cite{2002ApJ...575..484G}, 24:\cite{2002MNRAS.336L..49S}, 25:\cite{2003A&A...398L..29S}, 26:\cite{2003A&A...401..677G}, 27:\cite{2003A&A...403..929K}, 28:\cite{2003AJ....125..343L}, 29:\cite{2003AJ....126..975T}, 30:\cite{2003AJ....126.1526B}, 31:\cite{2003AJ....126.2421C}, 32:\cite{2003AJ....126.2487B}, 33:\cite{2003ApJ...586..512B}, 34:\cite{2003ApJ...592.1186B}, 35:\cite{2003IAUS..211..197W}, 36:\cite{2004A&A...416L..17K}, 37:\cite{2004A&A...425..519S}, 38:\cite{2004AJ....127.2948V}, 39:\cite{2004AJ....127.3553K}, 40:\cite{2004AJ....128.1733G}, 41:\cite{2004ApJ...604L..61C}, 42:\cite{2004ApJ...614L..73B}, 43:\cite{2005A&A...440..967S}, 44:\cite{2005A&A...440.1061L}, 45:\cite{2005AJ....129..511B}, 46:\cite{2005AJ....130.2326T}, 47:\cite{2005AN....326..974S}, 48:\cite{2006AJ....131.2722C}, 49:\cite{2006AJ....132..891R}, 50:\cite{2006AJ....132.2074M}, 51:\cite{2006ApJ...639.1095B}, 52:\cite{2006ApJ...639.1114R}, 53:\cite{2006ApJS..166..585B}, 54:\cite{2007A&A...466.1059K}, 55:\cite{2007A&A...474..653V}, 56:\cite{2007AJ....133..439C}, 57:\cite{2007AJ....134.1162L}, 58:\cite{2007MNRAS.378..901F}, 59:\cite{2008AJ....135..785W}, 60:\cite{2008AJ....136.1290R}, 61:\cite{2008ApJ...674..451B}, 62:\cite{2008ApJ...685.1183L}, 63:\cite{2008ApJ...689.1295K}, 64:\cite{2008MNRAS.384.1399J}, 65:\cite{2009AJ....137.3345C}, 66:\cite{2009ApJ...705.1416R}, 67:\cite{2010A&A...512A..37S}, 68:\cite{2010A&A...517A..53M}, 69:\cite{2010AJ....139.1808S}, 70:\cite{2010AJ....140..110G}, 71:\cite{2010ApJ...723..684B}, 72:\cite{2010MNRAS.409..552G}, 73:\cite{2011AJ....141...54A}, 74:\cite{2012ApJ...752...56F}, 75:\cite{2012ApJS..201...19D}, 76:\cite{2013AJ....146..161M}, 77:\cite{2013ApJ...772...79A}, 78:\cite{2013ApJS..205....6M}, 79:\cite{2013MNRAS.433.2054S}, 80:\cite{2013PASP..125..809T}, 81:\cite{2014A&A...565A..20S}, 82:\cite{2014AJ....147...34S}, 83:\cite{2014AJ....147...94D}, 84:\cite{2014ApJ...783..121G}, 85:\cite{2014ApJ...794..143B}, 86:\cite{2015ApJ...798...73G}, 87:\cite{2015ApJS..219...33G}, 88:\cite{2016AJ....152...24W}}

\end{table*}

To minimize differential reddening corrections \citep{1992AJ....103..638M} we
attempt to observe all targets within 30\,min of the meridian. The total time
for each target is 10-25\,min which enable an average of 3-4
targets/hour. Our observing runs were usually allocated in blocks of 3 nights
spread throughout the year.  Observations began on April 9th 2007 and using nights
obtained via Brazilian and ESO allocations continued for four years
until July 21 2011. After this date this telescope was no longer available
through ESO or Brazil and we obtained three additional runs in March 2014, October 2015
and February 2016 from the CNTAC, OPTICON and a few service observations on MPIA time. In
these extra runs we were able to re-observe most targets extending the
coverage to over 8 years providing important leverage to separate
parallax and proper motion components.

\subsection{Target selection}

The target lists had to meet a number of practical and scientific
considerations. The combination of a variable time allocation and the
requirement of observing objects close to the meridian required us to have a
target list that covered the whole 24 hour Right Ascension range
uniformly. To give flexibility for matching targets to conditions, and, to
ensure that any target was only observed in 2 out of each 3 night run, we
built redundancy into the list. With these requirements in mind we
adopted the following criteria:
 
\begin{enumerate}
\item Southern ($\delta < 0\degr$) confirmed L and T dwarfs discovered before
  April 2007,
\item Magnitude in the $z$ band brighter than 19,
\item Between 6-8 objects in any RA hour, 
\item The brightest examples within each spectral bin,
\item A uniform spectral class distribution for L dwarfs,
\item A photometric distance smaller than 50\,pc.
\end{enumerate}
The photometric distances were estimated using the 2MASS \cite[Two Micron All
  Sky Survey][]{2006AJ....131.1163S} magnitudes transformed to the MKO system
using \cite{2004PASP..116....9S} and the colour - absolute magnitude
compilation given in \cite{2004AJ....127.3553K}.  This produced an original
target list of 140 targets that can be found in \cite{2011AJ....141...54A}.

In Table~\ref{baseinfo} we list the \NumI\ targets published here including a
short name for each object used throughout this paper, the discovery name, the
$z$ band magnitude adopted at the beginning of the program and, when they
exist, published values of optical/NIR spectral types, radial velocities and
parallaxes. The last column summarizes any published indications of
multiplicity, e.g. if the object is an unresolved binary (at the nominal WFI
resolution), in a wide binary system or a moving group candidate. The
distribution of the \NumI\ targets is shown in Fig.~\ref{histrogram} with
respect to all L0 to T2 objects with published parallaxes.

\subsection{Image reduction procedure}
All images were bias corrected and flat fielded using standard {\it IRAF
  CCDPROC} procedures. The WFI $z$-band images have strong interference
fringes that were removed using {\it RMFRINGE } with a fringe map made with
three steps: 1)  mask out all objects in all short exposures and four of the
long exposures; 2)  make a median image of the unmasked pixels scaling all
images by the exposure time; 3) smooth the median image using a 5 pixel box
car average. After  subtracting this fringe map scaled by the exposure time
from the cleaned images we again make a new fringe map and again subtract it,
this time scaled by the mean sky count. We did not use all the long exposures
in the construction of the fringe map as the move-target-to-pixel and 
masking procedures are not perfect so the resultant fringe map using
all frames often had a  halo around the target position. The telescope
pointing is only good to a few arcseconds so in the location frames the
target is rarely in the same position and this halo problem does not occur.

In \cite{2011AJ....141...54A} we adopted the Torino Observatory Parallax
Program \citep[TOPP]{SMA99A} centroiding procedures but, as discussed in
\cite{2013AJ....146..161M}, we found the Cambridge Astronomy Survey Unit's 
\textit{imcore} maximum likelihood barycentre (CASUTOOLS, v 1.0.21) more
consistent so we have adopted that package to determine the centroids of all
objects in the field. 

\subsection{Astrometric parameter determination}
The astrometric reduction was carried out using TOPP pipeline procedures and
the reader is referred to \cite{SMA99A} for details, here we just outline the
main steps.  A base frame, observed on a night with good seeing, was selected
and the measured x,y positions of all objects were transformed to a standard
coordinate $\xi,\eta$ system determined from a gnomic projection of the {\em
  Gaia} DR1 objects in the frame. All subsequent frames were transformed to
this standard coordinate system with a simple six constant linear astrometric
fit using all common objects except the target.  We then removed any frames that had
an average reference star error larger than the mean error for all frames plus
three standard deviations about that mean in either coordinate, or, had less
than 12 stars in common with the base frame.

Since the target is not used in the fit, its positional change is a reflection
of its parallax and proper motion. We fit a simple 5 parameter model to this
positional change, and that of all the other objects in the field, to find
their astrometric parameters implicitly assuming all objects are single.  We
then iterate this procedure where, in addition to removing frames as described
above, we also remove stars with large errors over the sequence from the
objects used to astrometrically transform frames. Finally, for the target
solution we removed any observations where the combined residual in the two
coordinates is greater than three times the $\sigma$ of the whole solution.

The solutions were tested for robustness using bootstrap-like testing where we
iterate through the sequence selecting different frames as the base frame thus
making many solutions that incorporate varied sets of reference
stars and starting from different dates. We create the subset of all solutions
with: (i) a parallax within one $\sigma$ of the median solution; (ii) the number
of included observations in the top 10\%; and (iii) at least 12 reference
stars in common to all frames. From this subset, for this publication, we have
selected the one with the smallest error. More than 90\% of the solutions were
within one $\sigma$ of the published solution.

To the relative parallaxes we add a correction (COR in Table~\ref{results})
to find astrophysically useful absolute parallaxes. The COR is estimated
from the average magnitude of the common reference stars and the Galaxy model of
Mendez \& van Altena (1996\nocite{men96}) in the $z$ band. When \G~produces
proper motions and parallaxes of the anonymous field objects we will be able
to tie more precisely to the absolute system.

\section{Astrometric results}\label{Astrometricresults}
In Table~\ref{results} we present the astrometric results for the
\NumI\ targets, listed are positions at epoch 2010, epoch of the base frame,
parallaxes, proper motions, relative to absolute corrections applied (COR),
number of observations used in the final solution (N$_{*}$), number of
anonymous objects used as references in the transformation of sequence frames
to the base frame (N$_o$), epoch coverage of the frames in the final solution
($\Delta$T). The parallax errors range from \lowsig~to \highsig\,mas with a
median of \medsig\,mas.  We also constructed a comparison sample of objects
from the literature with parallaxes mainly from the list maintained by Trent
Dupuy\footnote{\url{www.as.utexas.edu/~tdupuy/plx/}}
\citep{2012ApJS..201...19D,2016ApJ...833...96L,2013Sci...341.1492D} adding \NumNP\ L0 to T8
objects.

\begin{table*}
 \caption{Astrometric Parameters of PARSEC targets.}
 \label{results}
 \begin{tabular}{ccccccccr}   
\\ 
\hline                                                
Target & $\alpha$   $\delta$ & Baseframe  & $\varpi_{abs} \pm \sigma $ & $\mu_{\alpha}\cos \delta \pm \sigma $  & $\mu_{\delta} \pm \sigma $  & COR & N$_{*}$,N$_o$,$\Delta$T \\
        &    deg. epoch 2010   & epoch    & mas               & mas/yr                      & mas/yr                     & mas     & yr        \\
\hline                                      

0004-4044   &     1.1477475,   -40.7392975 &    2009.56 &    77.48 $\pm$  4.64 &   668.67 $\pm$  1.30 & -1498.18 $\pm$  1.29 &   0.55 &   6,  26,   8.10 \\
0013-2235   &     3.4910061,   -22.5891701 &    2009.73 &    46.83 $\pm$ 11.55 &    60.45 $\pm$  6.66 &   -69.37 $\pm$ 11.22 &   0.44 &   8,   9,   3.21 \\
0016-4056   &     4.2488780,   -40.9482895 &    2007.77 &    44.68 $\pm$ 11.39 &   196.06 $\pm$  5.37 &    24.92 $\pm$  6.54 &   0.62 &   8,   5,   1.97 \\
0032-4405   &     8.2331843,   -44.0852233 &    2010.86 &    29.30 $\pm$  4.72 &   120.68 $\pm$  1.30 &   -95.88 $\pm$  1.38 &   0.52 &  10,  26,   8.00 \\
0034-0706   &     8.7374725,    -7.1008713 &    2009.97 &    55.83 $\pm$ 12.26 &   197.07 $\pm$  9.04 &  -160.68 $\pm$  6.22 &  -0.14 &   5,  12,   3.11 \\
0054-0031   &    13.5279543,    -0.5177292 &    2010.63 &    12.65 $\pm$  4.87 &   192.24 $\pm$  1.27 &  -157.02 $\pm$  1.64 &   0.69 &  10,  18,   8.11 \\
0058-0651   &    14.6777007,    -6.8570328 &    2008.64 &    32.95 $\pm$  4.77 &   143.15 $\pm$  1.01 &  -123.22 $\pm$  0.86 &   0.46 &   9,  30,   8.10 \\
0109-5100   &    17.2573429,   -51.0135480 &    2009.56 &    62.52 $\pm$  2.63 &   219.42 $\pm$  0.86 &    75.71 $\pm$  0.69 &   0.46 &  10,  29,   8.10 \\
0117-3403   &    19.4482615,   -34.0573392 &    2009.72 &    19.81 $\pm$  6.04 &    93.36 $\pm$  1.58 &   -45.84 $\pm$  2.17 &   0.25 &   6,  16,   8.09 \\
0128-5545   &    22.1098342,   -55.7592488 &    2007.67 &    50.24 $\pm$  5.96 &  -248.50 $\pm$  1.53 &   118.85 $\pm$  2.31 &   0.24 &   9,  24,   8.10 \\
0144-0716   &    26.1484149,    -7.2712842 &    2009.97 &    74.23 $\pm$  5.16 &   377.71 $\pm$  1.15 &  -187.14 $\pm$  1.39 &   0.11 &   8,   5,   8.10 \\
0147-4954   &    26.8864692,   -49.9140583 &    2009.96 &    25.54 $\pm$  2.99 &   -59.83 $\pm$  0.90 &  -265.91 $\pm$  0.70 &   0.63 &   9,  24,   8.09 \\
0205-1159   &    31.3736178,   -11.9914857 &    2009.96 &    54.09 $\pm$  3.90 &   429.44 $\pm$  0.93 &    52.87 $\pm$  0.89 &   0.45 &  10,  25,   8.10 \\
0219-1938   &    34.8675775,   -19.6452902 &    2010.86 &    32.62 $\pm$  4.92 &   182.41 $\pm$  1.76 &   -98.63 $\pm$  1.74 &   0.31 &   7,  13,   7.75 \\
0227-1624   &    36.7943950,   -16.4141398 &    2009.73 &    54.22 $\pm$  4.44 &   429.97 $\pm$  1.42 &  -300.66 $\pm$  1.16 &   0.37 &   7,  19,   8.09 \\
0230-0953   &    37.6879324,    -9.8849150 &    2007.67 &    30.44 $\pm$  2.78 &   150.08 $\pm$  0.65 &   -63.40 $\pm$  1.36 &   0.68 &   5,  24,   8.10 \\
0235-0849   &    38.9481393,    -8.8221609 &    2009.73 &    30.10 $\pm$  2.56 &   -50.62 $\pm$  1.79 &    17.37 $\pm$ 13.16 &   0.21 &   6,   7,   2.98 \\
0235-2331   &    39.0000280,   -23.5222961 &    2009.96 &    41.73 $\pm$  7.41 &    95.03 $\pm$  4.81 &    38.91 $\pm$ 10.53 &   1.07 &   8,  12,   2.97 \\
0239-1735   &    39.9274815,   -17.5961006 &    2008.82 &    29.71 $\pm$  2.93 &    55.57 $\pm$  0.67 &   -93.75 $\pm$  0.73 &   0.53 &   6,  22,   8.10 \\
0255-4700   &    43.7695401,   -47.0158453 &    2010.63 &   206.06 $\pm$  5.81 &  1012.52 $\pm$  2.13 &  -550.88 $\pm$  2.93 &   0.30 &   6,  15,   7.99 \\
0257-3105   &    44.3597881,   -31.0968375 &    2010.64 &   101.60 $\pm$  6.68 &   605.88 $\pm$  1.49 &   339.17 $\pm$  1.82 &   0.74 &   9,  12,   6.80 \\
0318-3421   &    49.7266300,   -34.3580202 &    2009.73 &    44.67 $\pm$ 15.60 &   392.91 $\pm$  2.86 &    47.07 $\pm$  2.89 &   0.36 &   8,   5,   6.81 \\
0357-0641   &    59.3404787,    -6.6904842 &    2010.65 &    10.70 $\pm$  4.14 &   140.54 $\pm$  0.83 &    10.90 $\pm$  0.96 &   0.39 &  11,  26,   8.10 \\
0357-4417   &    59.3626437,   -44.2918283 &    2010.64 &    16.77 $\pm$  2.99 &    64.18 $\pm$  0.60 &    -9.57 $\pm$  0.99 &   0.38 &  11,  25,   8.10 \\
0408-1450   &    62.1202147,   -14.8429450 &    2010.98 &    46.88 $\pm$  3.33 &   199.95 $\pm$  0.80 &   -97.45 $\pm$  1.06 &   0.55 &   9,  29,   8.10 \\
0423-0414   &    65.9515200,    -4.2340809 &    2008.96 &    71.97 $\pm$  3.23 &  -322.98 $\pm$  0.80 &    85.43 $\pm$  1.06 &   0.58 &   8,  27,   8.01 \\
0439-2353   &    69.7642946,   -23.8867150 &    2011.13 &    82.72 $\pm$  4.03 &  -112.81 $\pm$  1.14 &  -155.28 $\pm$  0.95 &   0.49 &  16,  35,   8.01 \\
0518-2828   &    79.7496314,   -28.4778602 &    2010.87 &    44.61 $\pm$  5.08 &   -75.57 $\pm$  1.32 &  -269.53 $\pm$  2.01 &   0.54 &  17,  26,   8.41 \\
0523-1403   &    80.9095684,   -14.0501604 &    2015.77 &    79.44 $\pm$  2.12 &   105.05 $\pm$  0.61 &   164.79 $\pm$  1.02 &   0.54 &  18,  36,   8.02 \\
0539-0059   &    84.9670678,    -0.9828936 &    2010.98 &    79.89 $\pm$  1.33 &   161.03 $\pm$  0.37 &   323.12 $\pm$  0.42 &   0.57 &  23,  39,   8.40 \\
0559-1404   &    89.8315807,   -14.0813086 &    2009.96 &    97.88 $\pm$  1.78 &   570.52 $\pm$  0.51 &  -339.63 $\pm$  0.69 &   0.58 &  28,  39,   8.02 \\
0614-2019   &    93.5502757,   -20.3226490 &    2011.13 &    35.32 $\pm$  2.23 &   140.92 $\pm$  0.79 &  -308.66 $\pm$  0.49 &   0.59 &  40,  45,   8.40 \\
0624-4521   &    96.1918822,   -45.3641483 &    2009.73 &    82.16 $\pm$  1.88 &   -34.18 $\pm$  0.62 &   368.76 $\pm$  0.93 &   0.59 &   8,  33,   8.02 \\
0639-7418   &    99.9833256,   -74.3123259 &    2010.86 &    50.78 $\pm$  7.54 &    18.36 $\pm$  2.25 &     0.93 $\pm$  2.33 &   0.58 &  16,  32,   8.89 \\
0641-4322   &   100.3275249,   -43.3740297 &    2010.98 &    50.72 $\pm$  1.18 &   211.86 $\pm$  0.38 &   625.75 $\pm$  0.60 &   0.58 &  30,  40,   8.41 \\
0719-5051   &   109.8835887,   -50.8615515 &    2011.23 &    33.35 $\pm$  1.42 &   174.36 $\pm$  0.42 &   -50.70 $\pm$  0.68 &   0.56 &  23,  47,   8.90 \\
0729-7843   &   112.2928751,   -78.7261905 &    2010.97 &    10.07 $\pm$  2.75 &  -152.88 $\pm$  0.92 &   137.06 $\pm$  1.13 &   0.54 &  28,  48,   8.89 \\
0828-1309   &   127.1413758,   -13.1564006 &    2010.87 &    84.24 $\pm$  1.40 &  -569.63 $\pm$  0.37 &     4.47 $\pm$  0.49 &   0.60 &  20,  43,   8.90 \\
0832-0128   &   128.0192365,    -1.4766709 &    2010.98 &    42.57 $\pm$  1.90 &    64.72 $\pm$  0.50 &    11.55 $\pm$  0.63 &   0.59 &  13,  46,   8.16 \\
0835-0819   &   128.9229611,    -8.3219362 &    2008.97 &   146.19 $\pm$  2.82 &  -559.34 $\pm$  0.66 &   309.40 $\pm$  0.47 &   0.60 &  32,  35,   8.90 \\
0859-1949   &   134.8551364,   -19.8244429 &    2011.21 &    71.22 $\pm$  3.54 &  -323.03 $\pm$  0.76 &   -97.72 $\pm$  0.68 &   0.57 &  37,  39,   8.89 \\
0909-0658   &   137.4890435,    -6.9720179 &    2009.23 &    35.99 $\pm$  2.19 &  -184.43 $\pm$  0.61 &    20.19 $\pm$  0.52 &   0.55 &  17,  39,   8.90 \\
0921-2104   &   140.3094109,   -21.0816119 &    2011.21 &    77.87 $\pm$  1.60 &   254.13 $\pm$  0.35 &  -915.10 $\pm$  0.45 &   0.56 &  18,  38,   8.90 \\
0922-8010   &   140.5818017,   -80.1766907 &    2009.24 &    40.29 $\pm$  4.35 &    39.40 $\pm$  1.19 &   -55.85 $\pm$  1.57 &   0.53 &  31,  34,   8.89 \\
0928-1603   &   142.1649531,   -16.0534848 &    2008.26 &    32.38 $\pm$  2.78 &  -157.34 $\pm$  0.59 &    26.36 $\pm$  0.72 &   0.56 &  20,  38,   8.90 \\
1004-1318   &   151.1675640,   -13.3058472 &    2009.96 &    37.89 $\pm$  1.92 &  -121.80 $\pm$  0.50 &  -190.37 $\pm$  0.55 &   0.48 &  13,  44,   8.90 \\
1004-3335   &   151.1625165,   -33.5869592 &    2014.22 &    45.80 $\pm$  2.87 &   343.58 $\pm$  0.58 &  -345.45 $\pm$  0.67 &   0.56 &  45,  31,   8.90 \\
1018-2909   &   154.7438393,   -29.1651446 &    2014.22 &    33.86 $\pm$  1.43 &  -342.53 $\pm$  0.37 &   -92.23 $\pm$  0.69 &   0.57 &  20,  40,   8.90 \\
1045-0149   &   161.3485657,    -1.8323859 &    2009.17 &    52.41 $\pm$  3.21 &  -488.66 $\pm$  0.70 &    -5.73 $\pm$  0.66 &   0.47 &  10,  29,   8.90 \\
1047-1815   &   161.8783338,   -18.2658328 &    2009.35 &    31.49 $\pm$  4.24 &  -352.60 $\pm$  0.80 &    43.65 $\pm$  0.63 &   0.57 &  12,  32,   8.89 \\
1058-1548   &   164.6985694,   -15.8047221 &    2009.17 &    49.22 $\pm$  3.11 &  -255.34 $\pm$  0.70 &    37.94 $\pm$  0.71 &   0.50 &  10,  38,   8.89 \\
1059-2113   &   164.9644957,   -21.2195072 &    2010.32 &    28.29 $\pm$  2.95 &   107.30 $\pm$  0.65 &  -160.43 $\pm$  0.65 &   0.40 &  12,  36,   8.89 \\
1122-3512   &   170.5339242,   -35.2109038 &    2011.12 &    78.61 $\pm$  5.96 &  -131.73 $\pm$  0.65 &  -263.52 $\pm$  1.34 &   0.60 &  22,  26,   8.01 \\
1122-3916   &   170.6511768,   -39.2687366 &    2014.21 &    32.49 $\pm$  7.62 &    49.75 $\pm$  1.11 &  -184.43 $\pm$  1.02 &   0.57 &  53,  26,   8.01 \\
1126-5003   &   171.6589674,   -50.0640251 &    2011.21 &    63.23 $\pm$  1.95 & -1583.13 $\pm$  0.56 &   454.42 $\pm$  0.40 &   0.42 &  66,  34,   7.92 \\
1154-3400   &   178.6753759,   -34.0108286 &    2011.21 &    30.15 $\pm$  3.16 &  -156.78 $\pm$  0.65 &    17.23 $\pm$  0.61 &   0.51 &  26,  35,   8.01 \\
1225-2739   &   186.4777455,   -27.6649768 &    2011.12 &    78.96 $\pm$ 11.41 &   374.84 $\pm$  1.33 &  -624.75 $\pm$  1.38 &   0.33 &  25,  21,   8.01 \\
1228-1547   &   187.0638399,   -15.7934534 &    2009.57 &    45.78 $\pm$  2.97 &   130.83 $\pm$  0.52 &  -179.44 $\pm$  1.13 &   0.50 &  15,  31,   8.01 \\
1246-3139   &   191.6235015,   -31.6594760 &    2009.37 &    88.38 $\pm$  3.43 &    -5.24 $\pm$  0.61 &  -560.84 $\pm$  0.96 &   0.39 &  37,  31,   8.01 \\
1254-0122   &   193.7232580,    -1.3796031 &    2010.46 &    68.04 $\pm$  6.53 &  -476.70 $\pm$  1.11 &   120.42 $\pm$  0.86 &   0.40 &   5,  28,   8.01 \\
1326-2729   &   201.5825064,   -27.4937769 &    2014.22 &    39.80 $\pm$  8.66 &  -362.07 $\pm$  1.22 &   -24.19 $\pm$  1.11 &   0.39 &  28,  25,   8.01 \\
 \hline
 \end{tabular}
\end{table*}
\begin{table*}
 \contcaption{Astrometric Parameters of PARSEC targets.}
 \begin{tabular}{ccccccccr}   
\\ 
\hline                                                
Target & $\alpha$   $\delta$ & Baseframe  & $\varpi_{abs} \pm \sigma $ & $\mu_{\alpha}\cos \delta \pm \sigma $  & $\mu_{\delta} \pm \sigma $  & COR & N$_{*}$,N$_o$,$\Delta$T \\
        &    deg. epoch 2010   & epoch    & mas               & mas/yr                      & mas/yr                     & mas     & yr        \\
\hline                                                
1331-0116   &   202.9526675,    -1.2837683 &    2016.17 &    49.98 $\pm$  4.59 &  -406.62 $\pm$  0.98 & -1035.16 $\pm$  1.54 &   0.26 &   8,  15,   7.91 \\
1341-3052   &   205.2984590,   -30.8811496 &    2011.21 &    26.64 $\pm$  4.08 &    19.13 $\pm$  0.77 &  -159.89 $\pm$  0.60 &   0.43 &  28,  28,   8.01 \\
1404-3159   &   211.2071517,   -31.9925574 &    2009.57 &    45.03 $\pm$  2.48 &   335.66 $\pm$  0.60 &   -19.96 $\pm$  1.01 &   0.43 &  47,  28,   8.01 \\
1425-3650   &   216.3656475,   -36.8411210 &    2014.21 &    79.64 $\pm$  3.60 &  -278.39 $\pm$  0.75 &  -466.42 $\pm$  1.16 &   0.41 &  34,  25,   8.89 \\
1438-1309   &   219.7296481,   -13.1530385 &    2011.22 &    30.65 $\pm$  3.06 &   147.37 $\pm$  0.79 &   -42.26 $\pm$  0.88 &   0.21 &  28,  31,   8.89 \\
1441-0945   &   220.4042039,    -9.7664631 &    2008.65 &    25.83 $\pm$  2.74 &  -193.91 $\pm$  0.77 &   -12.71 $\pm$  0.50 &   0.43 &  16,  28,   8.90 \\
1457-2121   &   224.3159237,   -21.3687893 &    2014.21 &   162.09 $\pm$  9.54 &  1034.15 $\pm$  2.71 & -1694.83 $\pm$  2.51 &   0.31 &  30,  30,   8.89 \\
1507-1627   &   226.9482001,   -16.4636115 &    2011.23 &   135.86 $\pm$  1.59 &  -145.81 $\pm$  0.37 &  -890.96 $\pm$  0.50 &   0.35 &  31,  29,   8.90 \\
1520-4422B   &   230.0065948,   -44.3786651 &    2008.26 &    48.50 $\pm$  3.09 &  -625.08 $\pm$  0.64 &  -390.15 $\pm$  1.20 &   0.35 & 210,  33,   8.90 \\
1523-2347   &   230.7768454,   -23.7980225 &    2007.27 &    29.06 $\pm$  1.58 &  -157.48 $\pm$  0.39 &   -15.95 $\pm$  0.57 &   0.32 &  20,  28,   8.90 \\
1530-8145   &   232.6076191,   -81.7608821 &    2014.21 &     8.34 $\pm$  2.22 &  -578.30 $\pm$  0.48 &  -278.02 $\pm$  1.33 &   0.50 &  43,  24,   8.01 \\
1534-2952   &   233.7078856,   -29.8751170 &    2008.26 &    60.27 $\pm$  3.11 &    98.39 $\pm$  0.63 &  -258.96 $\pm$  0.82 &   0.39 &  82,  28,   8.90 \\
1539-0520   &   234.9263159,    -5.3449277 &    2009.56 &    60.03 $\pm$  2.22 &   599.36 $\pm$  0.79 &   107.77 $\pm$  0.54 &   0.47 &  17,  38,   8.02 \\
1547-2423   &   236.9461995,   -24.3974581 &    2008.16 &    25.51 $\pm$  1.80 &  -139.11 $\pm$  0.60 &  -129.20 $\pm$  0.54 &   0.40 &  34,  37,   8.90 \\
1548-1636   &   237.2423980,   -16.6009345 &    2011.23 &    35.58 $\pm$  2.10 &  -199.25 $\pm$  0.59 &  -116.78 $\pm$  0.70 &   0.41 &  15,  31,   8.90 \\
1618-1321   &   244.6873677,   -13.3586183 &    2014.22 &    22.59 $\pm$  1.87 &  -105.68 $\pm$  0.54 &   -79.15 $\pm$  0.74 &   0.35 &  22,  33,   8.90 \\
1620-0416   &   245.1078344,    -4.2755193 &    2014.21 &    35.58 $\pm$ 12.97 &  -415.26 $\pm$  1.17 &   -11.38 $\pm$  1.02 &   0.28 &  30,  16,   8.90 \\
1633-0640   &   248.4964332,    -6.6826880 &    2011.22 &    35.67 $\pm$  4.08 &  -274.45 $\pm$  1.61 &  -230.01 $\pm$  1.61 &   0.28 &  60,  24,   8.89 \\
1636-0034   &   249.0022053,    -0.5819366 &    2009.24 &    25.84 $\pm$  2.81 &  -360.80 $\pm$  1.73 &  -209.72 $\pm$  1.82 &   0.35 &  28,  22,   3.95 \\
1645-1319   &   251.3409440,   -13.3332700 &    2009.38 &    88.28 $\pm$  1.30 &  -357.22 $\pm$  0.60 &  -800.00 $\pm$  1.14 &   0.33 &  51,  31,   8.90 \\
1705-0516   &   256.4518081,    -5.2797930 &    2011.22 &    53.30 $\pm$  1.26 &   120.84 $\pm$  0.55 &  -115.79 $\pm$  0.44 &   0.33 &  24,  37,   6.94 \\
1707-0558   &   256.8478553,    -5.9736454 &    2010.63 &    89.26 $\pm$  2.20 &    91.55 $\pm$  0.76 &    18.69 $\pm$  0.60 &   0.33 &  72,  31,   8.90 \\
1750-0016   &   267.6022252,    -0.2702677 &    2014.21 &   110.43 $\pm$  1.08 &  -399.46 $\pm$  0.55 &   202.81 $\pm$  0.69 &   0.30 &  62,  38,   6.94 \\
1753-6559   &   268.4379566,   -65.9997520 &    2008.26 &    65.74 $\pm$  2.67 &   -50.27 $\pm$  1.51 &  -334.23 $\pm$  1.13 &   0.47 &  64,  47,   6.94 \\
1828-4849   &   277.1497034,   -48.8177596 &    2009.72 &    78.44 $\pm$  4.87 &   229.11 $\pm$  1.71 &    86.09 $\pm$  2.50 &   0.36 & 104,  23,   8.48 \\
1840-5631   &   280.0788728,   -56.5210301 &    2007.27 &    11.02 $\pm$  4.66 &   -84.46 $\pm$  2.65 &  -166.76 $\pm$  7.15 &   0.40 &  58,  27,   4.28 \\
1928-4356   &   292.2163563,   -43.9409577 &    2010.60 &    35.06 $\pm$  4.63 &    66.34 $\pm$  1.52 &  -284.68 $\pm$  2.71 &   0.29 &  44,  34,   8.10 \\
1936-5502   &   294.0088562,   -55.0430750 &    2009.33 &    46.82 $\pm$  2.05 &   202.39 $\pm$  1.25 &  -287.87 $\pm$  2.37 &   0.31 &  30,  38,   3.89 \\
1956-1754   &   299.0643263,   -17.9070956 &    2009.73 &    25.82 $\pm$  1.58 &     6.18 $\pm$  0.65 &   -27.52 $\pm$  0.79 &   0.38 &  42,  38,   8.10 \\
2002-0521   &   300.7107882,    -5.3650469 &    2010.59 &    59.76 $\pm$  7.30 &  -120.39 $\pm$  2.63 &  -107.71 $\pm$  3.42 &   0.46 &  72,  32,   8.10 \\
2011-6201   &   302.9870707,   -62.0212510 &    2009.72 &    11.45 $\pm$  3.98 &   314.94 $\pm$  2.23 &  -394.10 $\pm$  2.25 &   0.34 &  33,  26,   3.88 \\
2023-5946   &   305.8693574,   -59.7811637 &    2011.55 &    10.89 $\pm$  5.01 &    70.67 $\pm$  3.36 &   -14.38 $\pm$  1.77 &   0.39 &  23,  25,   3.88 \\
2026-2943   &   306.5661130,   -29.7212734 &    2015.76 &    35.29 $\pm$  7.00 &    17.84 $\pm$  2.10 &  -355.40 $\pm$  2.43 &   0.23 &  15,  27,   8.10 \\
2041-3506   &   310.4286404,   -35.1127254 &    2010.63 &    22.91 $\pm$  3.45 &    40.39 $\pm$  1.13 &  -134.48 $\pm$  1.42 &   0.34 &  28,  24,   8.09 \\
2045-6332   &   311.2604376,   -63.5357349 &    2008.65 &    50.55 $\pm$  2.80 &    75.33 $\pm$  1.10 &  -204.59 $\pm$  0.99 &   0.46 &  17,  28,   8.10 \\
2057-0252   &   314.4753838,    -2.8753406 &    2010.33 &    70.40 $\pm$  2.70 &     3.13 $\pm$  0.89 &   -90.84 $\pm$  0.89 &   0.52 &  14,  31,   8.10 \\
2101-2944   &   315.4683482,   -29.7347644 &    2008.64 &    30.78 $\pm$  5.81 &    65.71 $\pm$  4.32 &     3.09 $\pm$ 17.68 &   0.26 &  12,   5,   1.95 \\
2104-1037   &   316.0640199,   -10.6278714 &    2010.59 &    55.89 $\pm$  4.42 &   596.79 $\pm$  2.40 &  -284.53 $\pm$  2.69 &   0.34 &  10,   7,   2.92 \\
2107-4544   &   316.9757837,   -45.7351717 &    2009.73 &    34.88 $\pm$  7.03 &    91.34 $\pm$  2.80 &   -15.52 $\pm$  2.63 &   0.46 &  12,  26,   8.10 \\
2130-0845   &   322.6869987,    -8.7558442 &    2009.38 &    36.98 $\pm$  9.48 &   350.92 $\pm$  2.38 &   -30.69 $\pm$  1.19 &   0.47 &  11,  23,   8.10 \\
2132-1452   &   323.2037671,   -14.8823207 &    2009.73 &    31.53 $\pm$ 12.63 &  -100.62 $\pm$  5.43 &  -145.93 $\pm$  7.32 &   0.30 &  24,  23,   3.88 \\
2150-7520   &   327.5755503,   -75.3442968 &    2010.46 &    42.67 $\pm$  8.28 &   881.96 $\pm$  2.16 &  -297.71 $\pm$  1.92 &   0.42 &  12,  22,   8.10 \\
2157-5534   &   329.4546555,   -55.5783892 &    2010.46 &    35.12 $\pm$  3.85 &    43.92 $\pm$  0.98 &   -10.75 $\pm$  1.06 &   0.36 &   9,  27,   8.10 \\
2158-1550   &   329.5190934,   -15.8362139 &    2009.38 &    44.69 $\pm$  7.32 &    39.20 $\pm$  2.07 &   -57.36 $\pm$  3.23 &   0.24 &  17,  13,   3.88 \\
2204-5646   &   331.0642768,   -56.7897998 &    2007.67 &   277.37 $\pm$ 14.60 &  4004.97 $\pm$  9.65 & -2575.05 $\pm$  9.02 &   0.39 &  18,  13,   3.88 \\
2206-4217   &   331.6879882,   -42.2897395 &    2009.56 &    36.79 $\pm$ 14.87 &   126.88 $\pm$ 15.20 &  -176.59 $\pm$  8.13 &   0.37 &  13,  12,   2.93 \\
2209-2711   &   332.3409290,   -27.1928581 &    2015.76 &    54.13 $\pm$  9.20 &    -5.95 $\pm$  2.36 &  -122.17 $\pm$  2.72 &   0.26 &  17,  19,   8.09 \\
2213-2136   &   333.4372519,   -21.6024355 &    2007.67 &    41.68 $\pm$ 12.46 &     2.23 $\pm$  9.87 &   -52.05 $\pm$  5.21 &   0.30 &  11,   6,   1.89 \\
2224-0158   &   336.1840312,    -1.9838795 &    2007.67 &    83.69 $\pm$  4.06 &   471.41 $\pm$  0.85 &  -863.28 $\pm$  1.92 &   0.56 &  11,  13,   8.10 \\
2252-1730   &   343.0459795,   -17.5033484 &    2009.96 &    50.09 $\pm$  4.62 &   397.32 $\pm$  0.91 &   136.27 $\pm$  1.11 &   0.31 &   9,  24,   8.01 \\
2254-2840   &   343.7164432,   -28.6736748 &    2009.38 &    32.87 $\pm$  3.60 &    -3.48 $\pm$  2.80 &    31.90 $\pm$  2.27 &   0.21 &  14,   8,   2.06 \\
2255-0034   &   343.8710403,    -0.5765119 &    2011.55 &    20.20 $\pm$  6.28 &   -41.42 $\pm$  1.49 &  -176.89 $\pm$  1.86 &   0.28 &   9,  28,   8.09 \\
2310-1759   &   347.5770339,   -17.9868960 &    2009.73 &    24.75 $\pm$  4.79 &    32.85 $\pm$  1.19 &  -285.87 $\pm$  1.08 &   0.35 &   8,   9,   8.11 \\
2318-1301   &   349.7267431,   -13.0204057 &    2007.67 &    81.47 $\pm$ 13.99 &  -797.18 $\pm$  4.81 &  -255.13 $\pm$  6.07 &   0.20 &  10,  14,   6.03 \\
2330-0347   &   352.5948267,    -3.7885807 &    2009.96 &    42.38 $\pm$  9.51 &   203.44 $\pm$  1.92 &    12.48 $\pm$  1.21 &   0.19 &   8,  20,   8.11 \\
2346-5928   &   356.6121968,   -59.4783859 &    2010.86 &    15.44 $\pm$  3.28 &   251.08 $\pm$  0.97 &    56.34 $\pm$  0.83 &   0.50 &  11,  30,   8.10 \\
2351-2537   &   357.9614199,   -25.6263449 &    2009.73 &    41.21 $\pm$  4.45 &   341.29 $\pm$  1.10 &   198.30 $\pm$  1.16 &   0.36 &   7,  17,   8.10 \\

  \hline
 \end{tabular}
\end{table*}

\subsection{Comparison to published parallaxes}
\label{ComparisontoPublishedParallaxes}
\begin{figure}
\begin{center}
\includegraphics[scale=0.45]{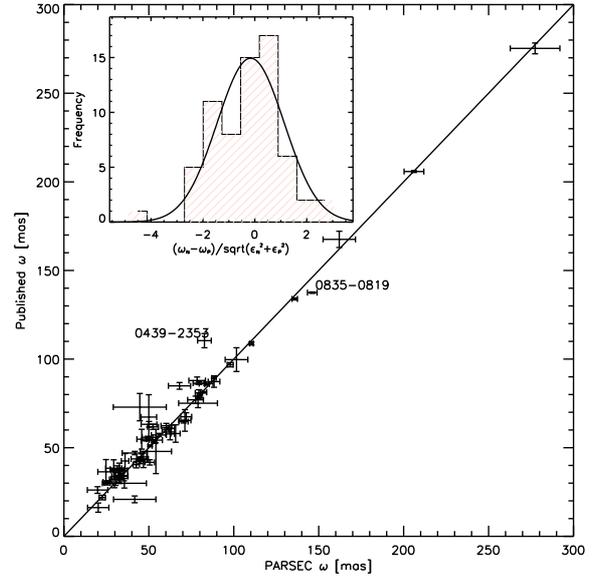}
\caption{Published vs PARSEC parallaxes. The solid line is the locus
  expected if the published and new values are equal. The two labeled objects
  have published and PARSEC parallaxes that differ by more than three times their
  combined errors and are discussed in
  Section~\ref{ComparisontoPublishedParallaxes}. The insert shows the
  distribution of the quantity $\frac{\varpi_{N} -
    \varpi_{P}}{\sqrt{\sigma_{N}^2 + \sigma_{p}^2 }}$ as discussed in Section~\ref{ComparisontoPublishedParallaxes}. }
\label{New_Published_pi}
\end{center}
\end{figure}

There are published estimates of the parallaxes for \NumP\ PARSEC targets of
which 24 were preliminary values from this program. The preliminary values
were found using different reduction procedures and different epoch spans,
hence we consider them as independent estimates. In
Fig.~\ref{New_Published_pi} we plot the \NumP\ published values listed in Table~\ref{baseinfo} versus results presented here from Table~\ref{results}.

Only two objects, 0439-2353 and 0835-0819, have parallaxes that differ by more
than three times the combined errors so warrant further consideration. For
0439-2353 \cite{2012ApJ...752...56F} find $110.4 \pm 4.0$\,mas while we obtain
$79.75 \pm 3.36$\,mas. The Faherty et al. observational coverage was 3 years
with 14 nights while we have 8 years and 18 nights.  We estimated the
photometric parallax to be (${\sim}80$\,mas) using the
\cite{2004AJ....127.2948V} absolute magnitude calibrations which is closer to
the PARSEC estimate. The difference between the Faherty et al. parallax and
the photometric parallax could be explained if the object is an un-resolved
binary but the residuals show no non-linear motion and a spectroscopic
investigation for binarity by \cite{2016MNRAS.455.1341M} also found no
evidence. We therefore consider our value more probable. For 0835-0819 we
obtain 146.19 $\pm$ 2.82 from 35 observations over 8.9 years while
\cite{2016AJ....152...24W} find 137.5 $\pm$ 0.4 using observations in 17
epochs over 6.2 years. The difference is just over three times the combined
error and the quoted error from \cite{2016AJ....152...24W} is very low
considering the observations were made on a similar system so we believe that
the larger than $3\sigma$ difference is probably because the errors are
underestimated.

For the \NumP\ PARSEC dwarfs with published parallaxes we calculated the
quantity
$\frac{\varpi_{N} - \varpi_{P}}{\sqrt{\sigma_{N}^2 + \sigma_{p}^2 }}$,
where $\varpi$ is the parallax, $\sigma$ the quoted errors and the subscripts $N$
and $P$ represent the new and published values respectively. If the measures
are unbiased and the errors are precise we expect this quantity to follow a
Gaussian distribution with a mean of zero and a standard deviation of one. For
the \NumP\ common objects the mean is -0.1 and the standard deviation is 1.3.
Applying the t-test at the 95\% level we find the mean is not significantly
different from zero, e.g. P(t)=0.06, while applying the F-test we find the
$\sigma$ is significantly different from one, e.g. P(F)=0.0001.  Since the
$\sigma$ is greater than one the implication is that the errors are
underestimated. The median PARSEC error is larger than the median published
error even though the programs were often very similar, however, without a
standard comparison it is difficult to isolate. The \G\ sample should allow a
complete characterisation of the errors of different procedures which can then
be applied to objects that are fainter than the \G\ limit in those programs.

\subsection{Comparisons within binaries}
Binary systems are a good test for parallax determinations and in particular
for the quality assurance of errors. Components in binary systems can be
considered to be at the same distance and if the system is a wide binary it is
appropriate to make independent solutions. In the PARSEC sample there are
\NumBin\ companions observable (e.g. in the field of view and not so bright
that they saturate) and in Table~\ref{bin} we reproduce the parallaxes and
proper motions of the PARSEC targets and the companions determined in this
program. No parallaxes or proper motions differ by more than $2\sigma$ between
the PARSEC values or the published values but this sample is too small to test
the precision of the error estimates.

\begin{table}
\begin{center}
 \caption{\label{bin} Wide binary systems in the PARSEC program}
\begin{tabular}{llll} 
\hline
Target     & $\varpi \pm \sigma $  &   $\mu_{\alpha} \pm \sigma$    & $\mu_{\delta} \pm \sigma$  \\
Companion  &  (mas)                &  (mas/yr)                    & (mas/yr)                 \\       
\hline
0004-4044$^{1,a}$  &   77.5 $\pm$ 4.6    &   668.7 $\pm$  1.3 & -1498.2 $\pm$  1.3 \\ 
GJ 1001A       &   86.4 $\pm$ 4.5    &   672.3 $\pm$  1.2 & -1500.1 $\pm$  1.4 \\  
         \noalign{\smallskip}
0235-2331       &     41.7 $\pm$  7.4 &    95.0 $\pm$  4.8 &    38.9 $\pm$ 10.5 \\
GJ 1048A$^{2,b}$ &     45.7 $\pm$  5.8 &    92.8 $\pm$  3.5 &   -18.1 $\pm$ 36.0\\ 
         \noalign{\smallskip}
1004-3335$^{c}$  &     45.8 $\pm$  2.9 &   343.6 $\pm$  0.6 &  -345.4 $\pm$  0.7\\ 
LHS 5166        &     49.8 $\pm$  2.2 &   349.6 $\pm$  0.4 &  -348.8 $\pm$  0.5 \\
         \noalign{\smallskip}
1441-0945$^{d}$  &     25.8 $\pm$  2.7 &  -193.9 $\pm$  0.8 &   -12.7 $\pm$  0.5\\ 
G 124-6         &     23.7 $\pm$  3.2 &  -203.5 $\pm$  0.8 &    -4.4 $\pm$  1.8 \\
\hline
\end{tabular}
\end{center}
\raggedright{Objects with published parallaxes 1:\,76.9 $\pm$4.0 \citep{2006AJ....132.2360H}, 2:\,47.2 $\pm$ 0.3 \citep{2016A&A...595A...2G}.}\newline
\raggedright{Reference for binarity. a:\,\cite{2004AJ....128.1733G}, b:\,\cite{2001AJ....122.1989W}, c:\,\cite{2005AN....326..974S}, d:\,\cite{2005A&A...440..967S}.}
\end{table}

\section{Photometric considerations}

\subsection{Internal photometry}
\begin{figure}
\centering
\includegraphics[width=8cm]{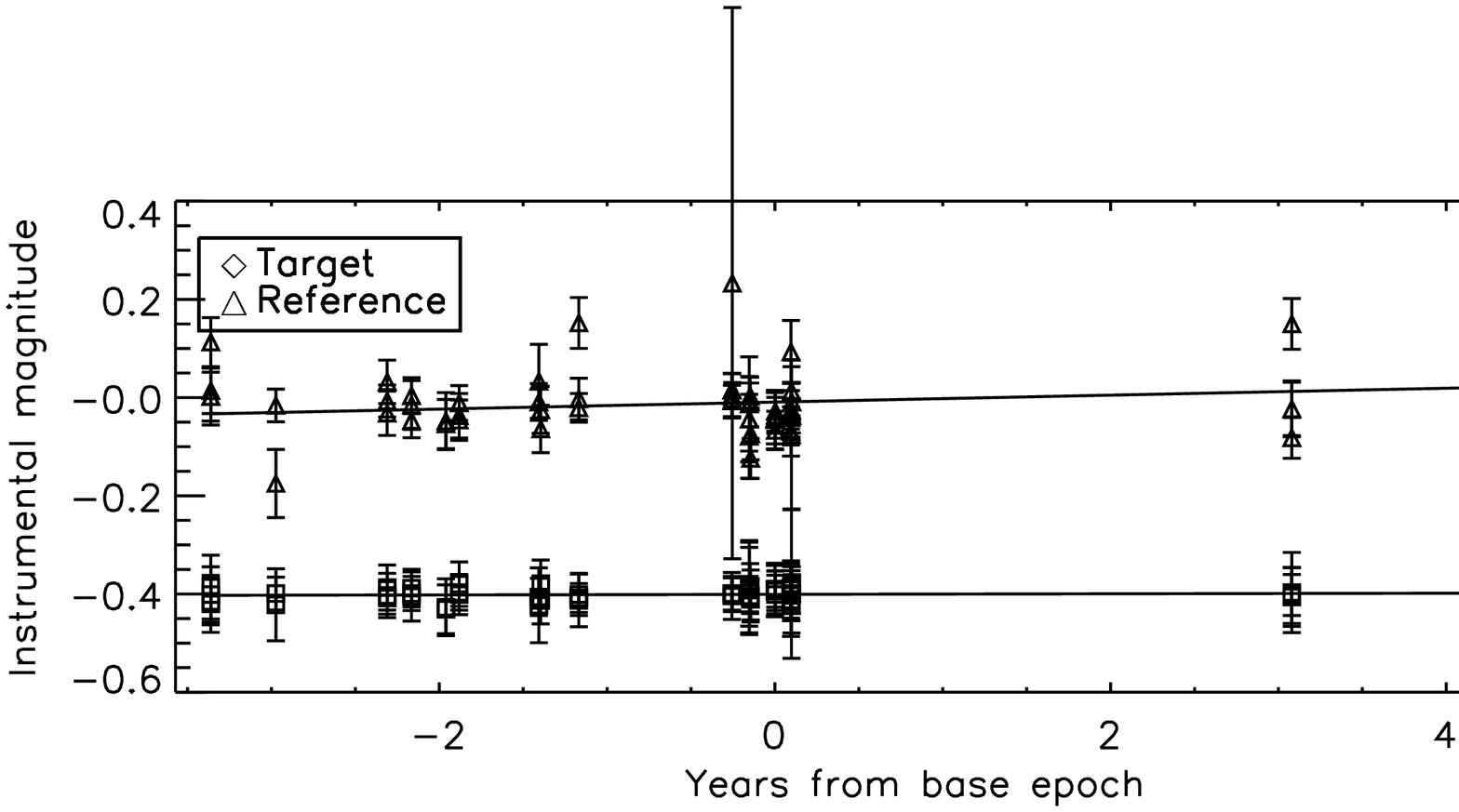}
\includegraphics[width=8cm]{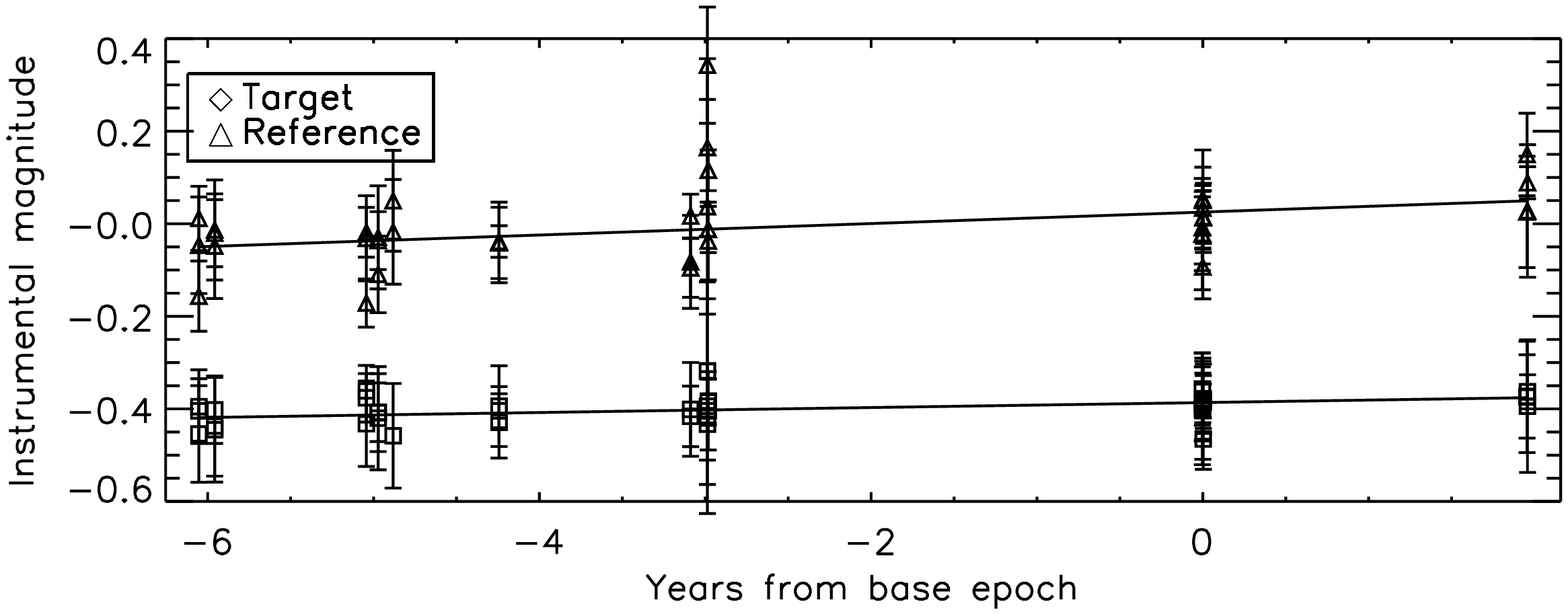}
\caption{ Instrumental $z$ magnitude variations as a function of time for
  0614-2019 (upper plot) and 1122-3916 (lower plot). As a comparison in each plot we have included an
  anonymous object that is nearby in position and magnitude space and included
  its variation offset by 0.4 magnitudes.}
\label{mag_res}
\end{figure}

During every observation we obtain precise relative photometry. The highest
observation frequency for our objects was monthly so the sampling is not
sufficient to find short period photometric variations but we can look for
long term variations. For each object we transformed the magnitudes to the
instrumental magnitude system of the base frame using the anonymous reference
stars with iterative 3$\sigma$ clipping and then found the linear correlation
between instrumental magnitudes and observation time of the target. The slopes
for most targets were within 3$\sigma$ of zero but two objects, 0614-2019 and
1122-3916, were found to have significant slopes of 0.0079 $\pm$ 0.0021 and
0.0124 $\pm$ 0.0037\,mag/yr respectively. For both objects the $\chi^2$ sum of
the two parameter fit was a statistically significant improvement over a one
parameter fit.  In Fig.~\ref{mag_res} we reproduce the magnitude variation of
these two objects over the observational sequence.  Possible explanations for
a long term photometric variation are discussed in \cite{2017MNRAS.468.3764S}
for Y dwarfs and in the K2 Ultracool Dwarfs Survey \citep[and references
  therein]{2018ApJ...861...76P} or the Weather on Other Worlds program
\citep[and references therein]{2017ApJ...840...83M} for L/T dwarfs.

\subsection{Literature photometry}
We compiled photometry on standard systems for the Full Sample 
from the large optical and infrared surveys: 
2MASS \citep{2006AJ....131.1163S};  
\G\ \citep{2016A&A...595A...1G}; 
Pan-STARRS \citep{2016arXiv161205560C}; 
SDSS \citep[Sloan Digital Sky Survey, ][]{2014ApJS..211...17A}; 
Visible and Infrared Survey Telescope for Astronomy (VISTA) VVV \citep[VISTA
  Variables in the Via Lactea, ][]{2010NewA...15..433M} -  
VMC \citep[VISTA Magellanic Survey, ][]{2011A&A...527A.116C} - 
VHS \citep[VISTA Hemisphere Survey, ][]{2013Msngr.154...35M} surveys; 
UKIDSS \citep[UKIRT Infrared Deep Sky Survey,][]{2007MNRAS.379.1599L}; 
and $WISE$ \citep{2014ApJS..211...17A}. 
Using these surveys it was possible to obtain
homogeneous magnitudes in bands ranging from Gunn G to $WISE$ W3 (the Gunn U
and $WISE$ W4 bands were not included as the number of objects with reliable
magnitudes was very low). The number of independent magnitude measures ranged
from  3 to 16 with a mean of 10 per target. 

For those objects with published parallaxes we took the weighted mean of the
PARSEC and published value with no outlier rejection. The complete dataset of
\NumFULL\ objects with photometry and parallaxes is available online here.

\subsection{Absolute magnitudes}
\label{sect:absmag}

\begin{figure}
\centering
\includegraphics[width=9cm]{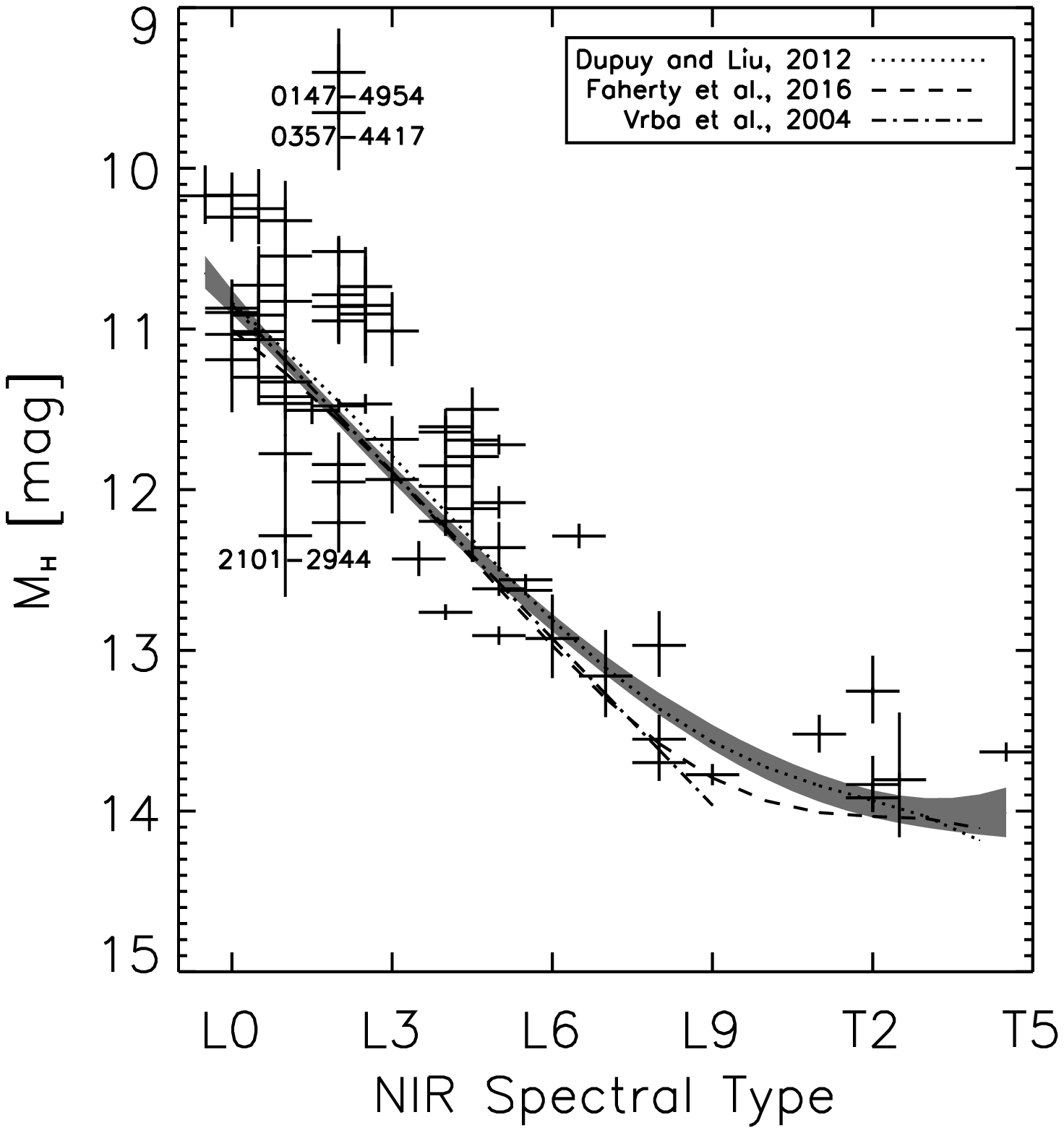}
\includegraphics[width=9cm]{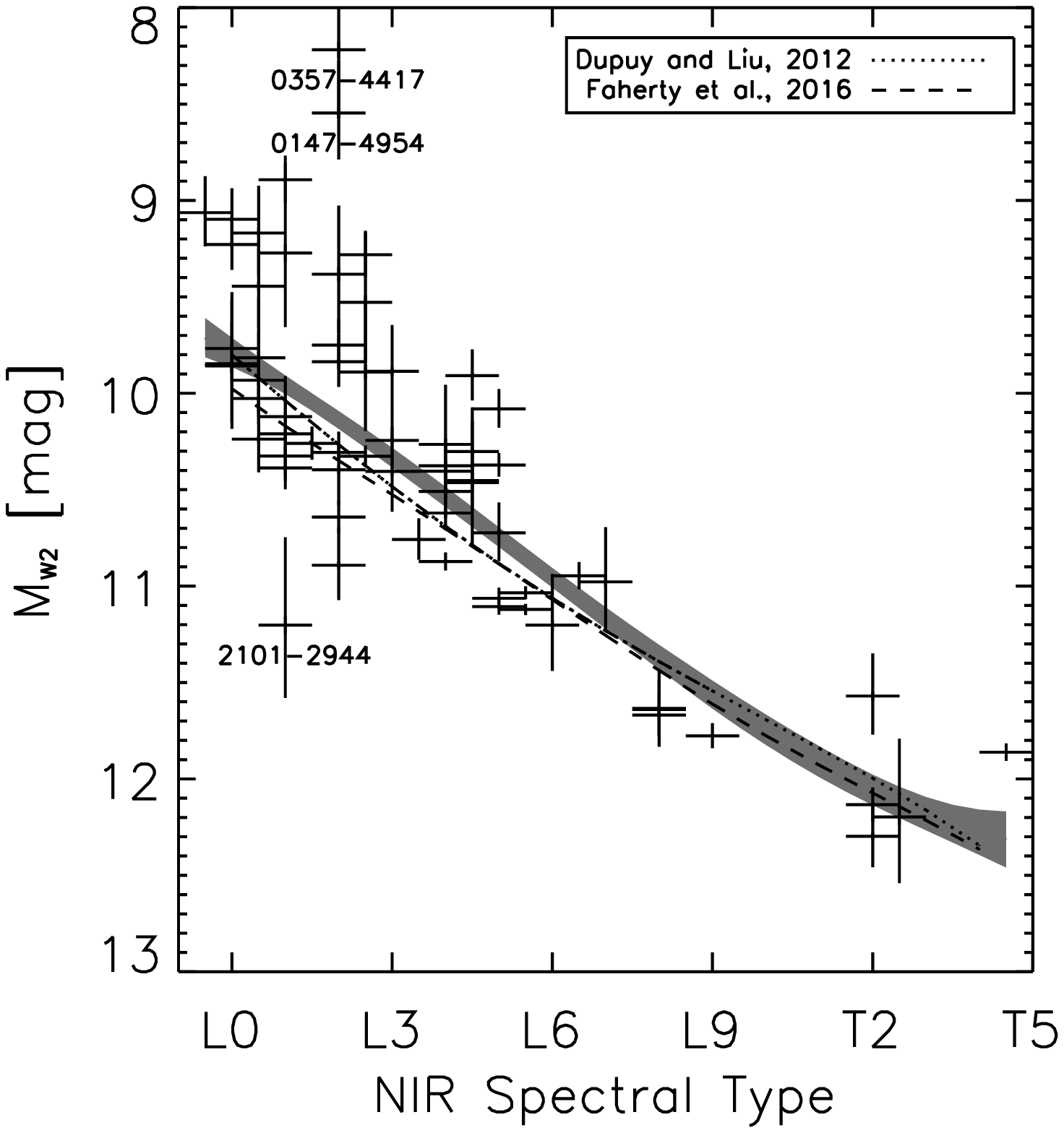}
\caption{PARSEC absolute magnitudes in the 2MASS $H$ (top) and $WISE~W2$
  (bottom) bands versus near-infrared spectral types. The length of the symbol
  arms represent the error which for spectral types is assumed to be 0.5
  types. The grey area represents one $\sigma$ confidence limits for a fit to
  published objects using literature parallaxes. We have included as
  comparisons the polynomial relations from Dupuy and Liu (2012), Faherty et
  al. (2016), and for $H$ only, Vrba et al. (2004) as indicated in the
  legend. }
\label{PublishedandNewTMASSJNIRSpectralType}
\end{figure}

\begin{table*}
\begin{center}
 \caption{\label{absmag} Absolute magnitude calibrations using Full Sample}
\begin{tabular}{lccccc} 
\hline
Passband  &   N  & $P1 $  & $ P2$   & $ P3$   & $\sigma$ \\
\hline
$r$    &   73 & -1.604e+02$\pm$ 8.648e+01 &  4.558e+00$\pm$ 2.375e+00 & -2.867e-02$\pm$ 1.630e-02 & 0.735\\ 
$i$    &   109 & -4.109e+01$\pm$ 2.947e+01 &  1.071e+00$\pm$ 7.969e-01 & -3.661e-03$\pm$ 5.382e-03 & 0.521\\
$z$    &   100 & -1.435e+02$\pm$ 2.123e+01 &  3.903e+00$\pm$ 5.698e-01 & -2.351e-02$\pm$ 3.819e-03 & 0.453\\
$y$    &   106 & -1.638e+02$\pm$ 2.111e+01 &  4.446e+00$\pm$ 5.664e-01 & -2.733e-02$\pm$ 3.796e-03 & 0.492\\
$Y$ MKO &   27 & -1.710e+02$\pm$ 3.006e+01 &  4.658e+00$\pm$ 7.970e-01 & -2.906e-02$\pm$ 5.271e-03 & 0.400\\
\hline\\  
$J$    &    170 & -1.530e+02$\pm$ 1.372e+01 &  4.129e+00$\pm$ 3.600e-01 & -2.543e-02$\pm$ 2.354e-03 & 0.558\\
$H$    &    167 & -8.511e+01$\pm$ 1.136e+01 &  2.317e+00$\pm$ 2.983e-01 & -1.353e-02$\pm$ 1.953e-03 & 0.456\\
$K_s$  &    160 & -4.320e+01$\pm$ 1.224e+01 &  1.181e+00$\pm$ 3.223e-01 & -5.964e-03$\pm$ 2.117e-03 & 0.427\\
$J$ MKO &   132 & -1.603e+02$\pm$ 1.730e+01 &  4.335e+00$\pm$ 4.539e-01 & -2.688e-02$\pm$ 2.969e-03 & 0.895\\
$H$ MKO &    40 & -1.116e+02$\pm$ 2.216e+01 &  3.017e+00$\pm$ 5.759e-01 & -1.811e-02$\pm$ 3.730e-03 & 0.438\\
$K$ MKO &    84 & -4.578e+01$\pm$ 1.474e+01 &  1.258e+00$\pm$ 3.871e-01 & -6.523e-03$\pm$ 2.534e-03 & 0.596\\
$W1$   &    162 &  2.264e+01$\pm$ 1.035e+01 & -5.415e-01$\pm$ 2.716e-01 &  5.167e-03$\pm$ 1.777e-03 & 0.415\\
$W2$   &    159 & -1.634e+01$\pm$ 1.118e+01 &  5.291e-01$\pm$ 2.932e-01 & -2.236e-03$\pm$ 1.917e-03 & 0.430\\
$W3$   &    114 & -4.693e+01$\pm$ 1.778e+01 &  1.394e+00$\pm$ 4.694e-01 & -8.431e-03$\pm$ 3.089e-03 & 0.507\\

\hline
\end{tabular}
\end{center}
\raggedright{Coefficients of polynomial fits to absolute magnitudes derived
  using parallaxes from Full Sample and large all sky photometry as described in Section~\ref{sect:absmag}.}
\end{table*}

In Fig.~\ref{PublishedandNewTMASSJNIRSpectralType} we plot the absolute
magnitudes of the PARSEC sample in the 2MASS $J$ and $WISE$ $W2$ bands vs
near-infrared spectral types. The crosses represent the PARSEC objects with
propagated errors in the magnitude axis and an assumed error of 0.5 types in
the spectral type axis. The grey area represents the one $\sigma$ confidence
limits of a second order fit to objects with published parallaxes. A fit to
just the PARSEC sample is within one sigma of fits to the Full Sample in all
magnitude bands. As a comparison we have plotted the polynomial relations
derived in the studies of \cite{2012ApJS..201...19D},
\cite{2016ApJS..225...10F} and for $H$ only \cite{2004AJ....127.2948V}.

In Table~\ref{absmag} we report the polynomial fits of the absolute magnitudes
to the published spectral types of the form:
\begin{equation}
  AbsMag = P1 + P2*Spt + P3*Spt^2 
\end{equation}
where $AbsMag$ is the absolute magnitude in the passband indicated in column 1
of Table~\ref{absmag}; $P1 ... P3$ are the parameters in columns 3-5 of
Table~\ref{absmag} and $SpT$ is the spectral type in numerical format,
e.g. L0=70, L1=71 ... T5=85. The absolute magnitude in the passbands $r$ to
$y$ refer to optical spectral types, the passbands $J$ to $W3$ refer to NIR
spectral types. For each passband we removed any objects which were tagged as
unresolved binaries or which had a $\varpi/\sigma < 5.$ and fitting was 
carried out using iterative outlier rejection of objects with residuals larger
than three times the overall fit error.

The labeled objects in Fig.~\ref{PublishedandNewTMASSJNIRSpectralType},
0147-4954, 0357-4417 and 2101-2944, are more than $3\sigma$ from the mean
absolute magnitude versus spectral type locus in at least two
pass-bands. 0147-4954 was included as a L2 but was reclassified as a M9 in
\cite{2013AJ....146..161M}, at this spectral type it is not over
luminous. 0357-4417 is a known unresolved binary \citep{2003AJ....126.1526B}
hence the brighter than average observed absolute magnitude. 2101-2944 is
under-luminous at the $3\sigma$ level in the $W1$ and 2MASS $K$ bands but shows no under-luminosity
in other bands and its spectra does not show any sign of peculiarity
\citep{2013AJ....146..161M} so the under-luminosity in these two bands is
unexplained.

\subsection{Spectral energy distribution analysis}
\label{SEDanalysis}
Using the multi-band photometry and our distances we are able to generate
spectral energy distributions (SEDs) for the PARSEC objects. While the SEDs
contain less information than the spectra the instrumental differences in
spectral observations render the SEDs globally more homogeneous. To compare
these to models we made use of the VO SED Analyzer \citep[VOSA,
][]{2008A&A...492..277B} which provides $\chi^2$ and Bayesian fitting to an
array of models and templates. For this data we adopted the $\chi^2$ fitting
to BT-Settl models \citep{2012RSPTA.370.2765A} with the following limits: $
700 < T_\mathrm{eff} < 4000$\,K, $ 3.5 < \log\,g < 5.5 $ and $-1 < \frac{\elem{Fe}}{\elem{H}} <
0.5$. We also limited the absorption in the $V$ band to 0.001 and turned off
the excess fitting option as the non-black body distribution of the spectra of
these objects was causing the excess fitting procedures to ignore the mid IR
magnitudes.

\begin{figure*}
\centering
\includegraphics[width=20cm]{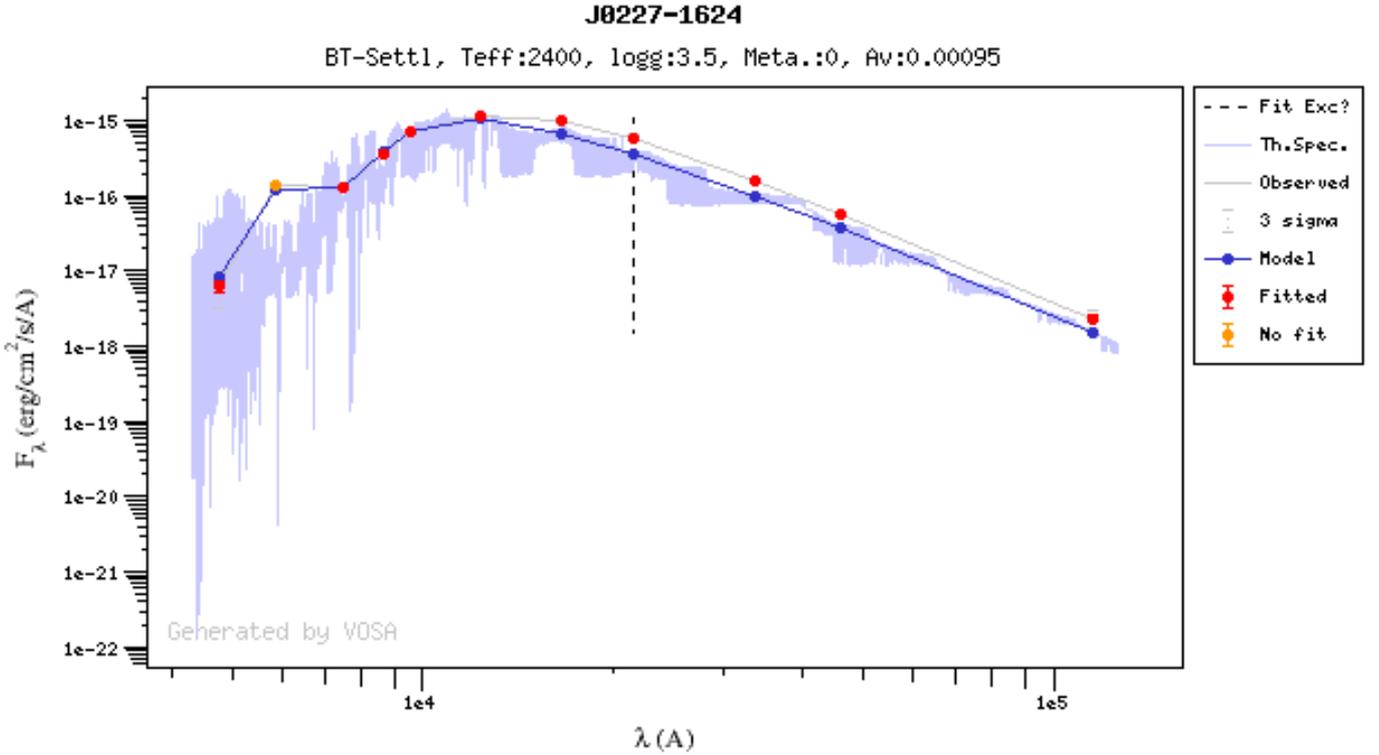}
\caption{ An example VOSA $\chi^2$ fitting of the 0227-1624 spectral energy
  distribution to BT-Settl models. The optical to $J$ bands appear to fit well
but the model underestimates the flux in the other near and mid-infrared bands
reasons for which are discussed in Section~\ref{SEDanalysis}.}
\label{0227-1624_VOSA.eps}
\end{figure*}

As an example in Fig.~\ref{0227-1624_VOSA.eps}  we plot the VOSA fit of
0227-1624 to the BT-Settl model spectra. 0227-1624 is a 16th $z$ band
magnitude L1/L0.5 object at 20\,pc with a 8\% error on the distance for which
we have magnitudes in 11 bands. The model slightly underestimates the near and
mid infrared bands but follows well for the optical bands. Reasons for this
over luminosity are beyond the scope of this paper as it will require a more
in depth study of the models and the fitting of VOSA.

The \G\ $G$ observation is plotted in orange because we manually removed it
from the fit. We find the nominal $G$ passband tends to over estimate the flux
of the L \& T dwarfs which can be seen in Fig.~\ref{0227-1624_VOSA.eps}.  This
systematic excess is also reflected in the model (the blue points) as this
uses the $G$ passband for the transformation, however, we felt this known
systematic error made using the \G\ magnitude for the \NumG\ objects in the
\G\ DR1 to constrain the fits premature.  The problem of the $G$ passband is
noted in the
\G\ documentation\footnote{\url{gaia.esac.esa.int/documentation/}},
and there is an empirical correction proposed\citep{2017A&A...608L...8M}.

\begin{figure}
\centering
\includegraphics[width=9cm]{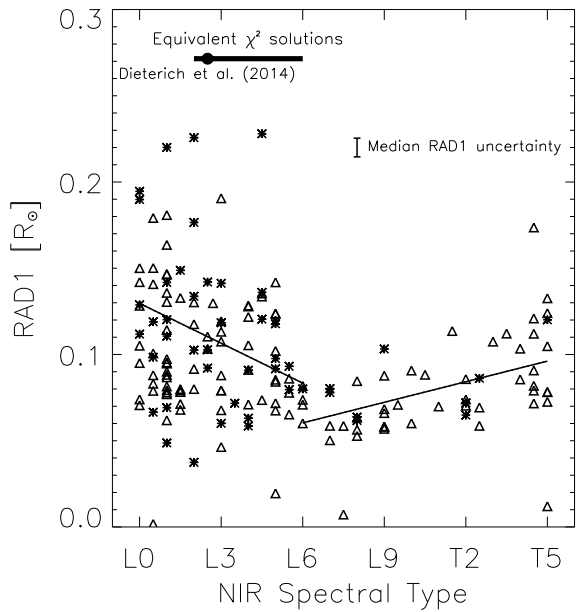}
\caption{Scaling radius (RAD1) in solar radii from VOSA BT-Settl model fitting
  vs near-infrared spectral types. The asterisks represent the 60 PARSEC
  objects and the diamonds the 132 literature objects that are not suspect
  binary or moving group members. The error bar in the top of the graph is the
  median error from VOSA for RAD1. The two lines represent two simple straight
  fits of RAD1 to Spectral Type on either side of L6. The black bar
  represents the range of solutions where the $\chi^2$ minimum did not vary
  significantly while the black circle is the minimum radius found by
  Dieterich et al. (2014). The sample and fitting
  procedure are discussed in Section~\ref{SEDanalysis}. }
\label{examinevosamodelRAD1vsNIR.eps}
\end{figure}

In Fig.~\ref{examinevosamodelRAD1vsNIR.eps} we plot the RAD1 radius from the
VOSA fits to BT-Settl models against the near-infrared spectral types. RAD1 is
the ``scaling radius'', i.e. the radius required to fit the observations to
the model based on the parallactic distance. From the PARSEC sample we removed
all objects which are known or suspect unresolved binary systems and known or
candidate moving group objects (see Section~\ref{MGsect}) leaving 60 objects.
We removed the moving group objects as these are in general young and we
wanted to be sure our sample was dominated by objects older than 1\,Gyr. In
Fig.~\ref{examinevosamodelRAD1vsNIR.eps} the targets plotted as diamonds are
literature objects and the PARSEC sample are plotted as asterisks.  The radius
of objects with ages greater than 1\,Gyr in this spectral range is predicted
to be a minimum at the spectral type that corresponds to the hydrogen burning
limit \citep[e.g.][]{2009AIPC.1094..102C,2011ApJ...736...47B}. Objects with earlier
spectral types than this limit are in hydrostatic equilibrium and the radius
decreases with the spectral type.  Objects with later spectral types than this
limit are degenerate and in this case the radius increases with the spectral
type. Hence, we expect to find a minimum RAD1 that corresponds to the spectral
type of objects at the hydrogen burning limit.

We experimented with a grid of trial spectral types fitting the RAD1 to the
spectral type on either side of the trial value with simple straight line
fits. In each fit we weight a common point given by the median around the
trial value to guide continuity. As an example in
Fig.~\ref{examinevosamodelRAD1vsNIR.eps} we have plotted the two line fit of
the observations assuming a minimum at L6. We then calculate the minimum
$\chi^2$ of the combined fits, which formally occurs if we choose the trial
value between spectral types L3 and L4, however there is no significant
difference between L2 to L6, the minimum is not well defined.  In
\cite{2014AJ....147...94D} for a smaller more refined sample they find the
position of the minimum at L2.5 which is the early end of our window. In
Dieterich et al. (2014) they use tailor made SED fitting that is calibrated
with radii of objects with radii measurements from interferometric
observations so we expect this to be a more robust estimate.

The larger sample we have included here is going to cover a range of ages,
metallicities and masses and as discussed in \cite{2011ApJ...736...47B} the
position of the minimum is dependent on age and metallicity so as this is a
mixed population we do not expect to have a unique clear minimum. In addition,
even for a given age the minimum may not be a single distinct value, for
example in the case of halo UCDs there is a narrow mass range in which
unsteady nuclear fusion occurs \citep{2017MNRAS.468..261Z}. If this occurs
even over a smaller range for younger UCDs it would result in spreading out of
the minimum. The general trend is of a steep dependence in the hydrogen burning
regime and a flatter change in the degenerate regime predicted by
\cite{2011ApJ...736...47B} is however confirmed.
 
%
%
%
\section{Kinematic considerations}

\subsection{Solar motion and velocity ellipsoids}

To calculate the $UVW$ velocity components in the galactic reference frame in
addition to position, proper motion and parallax we also require radial
velocities. We follow the general procedure of
\cite{1987AJ.....93..864J} except we use the transformation matrix from
equatorial coordinate to galactic coordinates taken from  the
introduction to the HIPPARCOS catalogue \citep{1997HIP...C......0E}.   
For the PARSEC sample \NumPRV\ objects have published radial velocities and
for the Full Sample \NumNPRV. For the objects without radial velocities we
estimate only the two components least impacted by assuming a zero radial
velocity following the recipe in \cite{2013AJ....145..102L}. From the Full
Sample of PARSEC and published L0-T8 dwarfs of \NumFULL\ objects we obtained
\Numu,\Numv,\Numw\ $U,V,W$ components respectively.

To isolate those dwarfs that are part of the thin disk population we use
$3\sigma$ clipping in each element and remove 34 objects from the Full Sample
of \NumFULL\ which includes 16 from the PARSEC subset. This is a simple
efficient method for the removal of non thin disk and outlier objects. In
Table~\ref{uvw} we report the mean and standard deviations of the velocity
components for the PARSEC, published and Full Sample after this cleaning.

The velocities are all heliocentric so the mean velocity vector indicates the
anti-motion of the Sun relative to this sample. From Table~\ref{uvw} we would
estimate the Solar motion compared to the Full Sample to be
$(U,V,W)_{\bigodot} = (5.4 \pm 1.4, 12.6\pm0.9, 7.6\pm1.1)$ km\,s$^{-1}$ which
can be compared directly to the Sun's velocity components inferred by larger
stellar groups, e.g. \cite{2010MNRAS.403.1829S}: $(U,V,W)_{\bigodot} =
(11.10^{+0.69}_{-0.75}, 12.24^{+0.47}_{-0.47}, 7.25^{+0.37}_{-0.36})$
km\,s$^{-1}$, and \cite{2009NewA...14..615F}: $(U,V,W)_{\bigodot} = (7.5 \pm
0.1,13.5 \pm 0.3, 6.8 \pm 0.1)$ km\,s$^{-1}$.  Considering the large
uncertainties, and given the agreement in the direction of rotation ($V$)
indicate that our sample, as a whole, is not moving very differently from the
local standard of rest.

The dispersions of the velocities ($\sigma_o$) are the result of the dynamical
evolution of our sample in the galactic disk. The dispersions derived here,
$(\sigma_{U}, \sigma_{V}, \sigma_{W}) = (24.4, 15.3, 14.2)$ km\,s$^{-1}$, are
consistent but generally smaller than other studies of L \& T dwarfs,
e.g. \cite{2007ApJ...666.1205Z}: (30.2, 16.5, 15.8)\,km s$^{-1}$,
\cite{2009AJ....137....1F}: (22,28,17)\,km s$^{-1}$, and
\cite{2010A&A...512A..37S}: (33.8,28.0,16.3)\,km s$^{-1}$.  The lower values
may be because of our stringent 3$\sigma$ clipping or the fact that our sample
are objects with parallaxes, which for a given spectral type are in general
brighter, and hence often younger.

\begin{table}
\begin{center}
 \caption{\label{uvw} Mean and dispersions of $UVW$ velocity components. $\nu$
   is the generic symbol used to indicate the components $U,V$ or $W$.}
\begin{tabular}{cccccc} 
\hline
Direction  & $\bar{\nu} \pm  \sigma_{\bar{\nu}} $        & $\sigma_{\nu}$ & N$_{\sigma}$ & $\sigma_{|W|\nu}$ & N$_{\sigma_{|W|\nu}}$ \\
 \&  Sample  &  km/s         &  km/s    &   &  km/s    &    \\ 
\hline\\
     $U$ PARSEC &  -5.3$\pm$2.5 &  22.6 &  85 &  22.5 &  29\\
 Not PARSEC &  -5.5$\pm$1.8 &  25.1 & 200 &  27.4 &  73\\
      Full  &  -5.4$\pm$1.4 &  24.4 & 285 &  26.1 & 102\\
\hline
     $V$ PARSEC & -12.8$\pm$1.7 &  15.6 &  88 &  22.4 &  32\\
 Not PARSEC & -12.5$\pm$1.0 &  15.2 & 220 &  19.9 &  93\\
       Full & -12.6$\pm$0.9 &  15.3 & 308 &  20.6 & 125\\
\hline
     $W$ PARSEC &  -7.4$\pm$2.1 &  14.4 &  46 &  12.4 &  46\\
 Not PARSEC &  -7.7$\pm$1.2 &  14.2 & 133 &  12.1 & 133\\
      All  &  -7.6$\pm$1.1 &  14.2 & 179 &  12.2 & 179\\
\hline
\end{tabular}
\end{center}
\end{table}

\subsection{Age estimation}
\cite{1977A&A....60..263W} finds that dynamical heating provides a
relation between observed velocity dispersion and mean age of a given stellar
population. Rearranging the velocity-dependent diffusion equation 13 from
Wielen for ages less than 3\,Gyr we get:
\begin{equation}\label{age1977}
   \tau = \frac{\tilde{\sigma}_{|W|\nu}^3(\tau)-\sigma_0^3}{1.5\gamma_{\nu}}
\end{equation}
where $\tau$ is the mean age, $\sigma_0=10$\,km/s and
$\gamma_{\nu}=1.4\times 10^{-5}$ (km/s)$^3$\,yr$^{-1}$ and $\tilde{\sigma}_{|W|\nu}$ is
total of the $|W|$ weighted velocity components, $\sigma_{|W|\nu}$ column in
Table~\ref{uvw}. These are calculated using: 
\begin{align}
 	\nonumber \sigma_{|W|U}^2=&\sum{|W|U^2}/\sum{|W|}\\
	\nonumber \sigma_{|W|V}^2=&\sum{|W|V^2}/\sum{|W|}\\
	\nonumber \sigma_{|W|W}^2=&0.5*\sum{|W|W^2}/\sum{|W|} 
\end{align}
  \noindent from Wielen (1977). 
The sum of the $|W|$ weighted velocities are 35.4\,km/s for
the PARSEC sub-sample and 34.1\,km/s for all objects corresponding
respectively to an age of 2.1 and 1.8\,Gyr. Using equation 16 from
Wielen for ages $>3$\,Gyr we also get an age of less than 3\,Gyr, so equation 13 is more appropriate.

These estimates of the age are younger than the ${\sim}5.1$\,Gyr in
\cite{2010A&A...512A..37S} and the $3.4\sim3.8$\,Gyr from
\cite{2015ApJS..220...18B} with similar samples and procedures, though both studies
benefitted from having radial velocities for all their target which we do not
have. Our estimates are however in agreement with $\tau =
1.7\pm0.3$\,Gyr found in \cite{2018PASP..130f4402W} using a similar
sample/procedure and also in \cite{2017ApJS..231...15D} where the median age
is 1.3\,Gyr for L dwarfs from dynamical masses and luminosities combined with
evolutionary models. The reason for this younger value may be, as before, a
result of our sample cleaning or because we are dominated by brighter
examples.  The \G\ results should be
complete to some limiting magnitude so it will be interesting to see what they reveal -- especially since the full
\G\ dataset will significantly constrain kinematically based ages.

\subsection{Moving group membership}
\label{MGsect}
We used the packages LocAting Constituent mEmbers In Nearby
Groups\footnote{\url{github.com/ariedel/lacewing}} \citep[hereafter
  LACEwING]{2017AJ....153...95R} and Bayesian Analysis for Nearby Young
AssociatioNs $\Sigma$\footnote{\url{github.com/jgagneastro/banyan_sigma_idl}}
\citep[hereafter BANYAN $\Sigma$]{2018ApJ...856...23G,2013ApJ...762...88M,2014ApJ...783..121G} to assess membership of
our sample to known moving groups starting from the assumption that they are
all field objects - e.g. that we have no spectral or colour evidence of youth.
This assumption is a conservative starting point as some objects are known to
have signs of youth and are often known moving group candidates as indicated
by {\it MG} in column 6 of Table~\ref{baseinfo}. Since we start from a
conservative position our moving group candidate indication should be more
robust and homogeneous.

The calculation of probability is different between LACEwING and BANYAN
$\Sigma$, in the first case the probabilities are considered independent while
in the second case the probabilities are required to sum up to 100\%,
i.e. the object is either in one of the included moving groups or it is a
field object. This generally leads to LACEwING having lower
probabilities. Based on a comparison of the objects that overlap we select as
candidate moving group members those with 80\% probability for BANYAN $\Sigma$
and 50\% for LACEwING.

We find \Nbaya\ objects from BANYAN $\Sigma$ and \Nlace\ from
LACEwING. The candidates that are not already published in the literature 
are listed in Table~\ref{MGnew} along with the name of the
moving group and the probability of membership.  Most of these candidates do
not have radial velocities, when these become available these probabilities 
should be revisited. Of the 13 PARSEC objects indicated in the literature
as moving group members six were not confirmed by either procedure:
0357-4417 \citep{2014ApJ...783..121G}, 1058-1548 \citep{2015ApJ...798...73G}, 1154-3400
\citep{2015ApJS..219...33G}, 1547-2423 \citep{2015ApJS..219...33G}, 1707-0558
\citep{2006AJ....132.2074M}, 2045-6332 \citep{2010MNRAS.409..552G}. For three of
these, 0357-4417, 1154-3400 and 1707-0558 the previous indication was 
made without a parallax which provides a strong new constraint.

\begin{table}
\begin{center}
 \caption{\label{MGnew} New moving group candidates with LACEwING selecting only
   probabilities greater than 50\% and BANYAN $\Sigma$ selecting only
   non-Field objects and probabilities greater than 80\%.}
\begin{tabular}{crr} 
\hline
PARSEC & BANYAN $\Sigma$ & LACEwING\\
Target & Group, Prob.  &   Group, Prob.   \\
\hline
   J0016-4056 & Field,     100         & $\beta$ Pictoris,      62      \\  
   J0032-4405 & AB Doradus,      94         & None                  \\  
   J0034-0706 & $\beta$ Pictoris,      85       & $\beta$ Pictoris,      63      \\  
   J0109-5100 & Field,     100         & $\beta$ Pictoris,      73      \\  
   J0117-3403 & Tucana-Horologium,      89       & None                  \\  
   J0205-1159 & Carina-Near,      95   & Hyades,      71       \\  
   J0219-1938 & Columba,      99       & AB Doradus,      80        \\  
   J0230-0953 & $\beta$ Pictoris,      96       & AB Doradus,      41        \\  
   J0408-1450 & $\beta$ Pictoris,      72       & AB Doradus,      72        \\  
   J0559-1404 & Field,     100         & AB Doradus,      64        \\  
   J0624-4521 & Field,      97         & Argus,      62        \\  
   J0859-1949 & $\beta$ Pictoris,      84       & Argus,      67        \\  
   J0928-1603 & Carina-Near,      92   & Carina-Near,      53  \\  
   J1018-2909 & $\beta$ Pictoris,      86       & None                  \\  
   J1045-0149 & Carina-Near,      97   & None                  \\  
   J1047-1815 & Carina-Near,      97   & None                  \\  
   J1126-5003 & Carina-Near,      94   & None                  \\  
   J1326-2729 & Carina-Near,      81   & None                  \\  
   J1425-3650 & AB Doradus,     100         & None                  \\  
   J1520-4422 & Carina-Near,      93   & None                  \\  
   J1753-6559 & AB Doradus,      92         & Argus,      95        \\  
   J1928-4356 & AB Doradus,      98         & AB Doradus,      54        \\  
   J1936-5502 & Field,     100         & AB Doradus,      60        \\  
   J2026-2943 & Field,     100         & AB Doradus,      55        \\  
   J2130-0845 & Carina-Near,      85   & AB Doradus,      79        \\  
   J2204-5646 & Carina-Near,      98   & None                  \\  
   J2206-4217 & AB Doradus,      96         & None                  \\  
   J2255-0034 & Field,     100         & AB Doradus,      93        \\  
   J2351-2537 & Field,     100         & Hyades,      84       \\  

\hline
\end{tabular}
\end{center}
\end{table}

%

\section{Conclusions}
\label{conclusion}

We have presented parallaxes and proper motions for \NumI\ objects. Using this
new sample we have examined their photometric and kinematic properties. In the
PARSEC sample we have identified candidate moving group members, found objects
with long term photometric variations, estimated the age of a local sample of
L \& T dwarfs and confirmed global trends of the predicted radius versus
spectral type variations.  We expect the number of sub-stellar objects with
known parallaxes to grow and the availability of statistically significant
samples will allow us to strengthen the constraints on models and to search
for fundamental calibrators.

Concurrent with this contribution there will be the second data release of the
\G\ mission which will have parallaxes and proper motions for 1.3 billion
sources and positions for a further 200
million\footnote{\url{www.cosmos.esa.int/web/gaia/dr2}}. For objects later
than L0 the number in the \G\ results will be quite modest, 500-1500 L0 to L5
dwarfs and 100-300 L5 to L9 dwarfs and less than 10 T dwarfs
\citep{2013A&A...550A..44S,2017MNRAS.469..401S,2018ApJ...862..173T}. In the
first \G\ data release only \NumG\ PARSEC objects were found and we do not
expect there to be many more with parallaxes and proper motions in the second
data release.

The PARSEC objects are at the magnitude limit of \G\ $G=21.3$
\citep{2016A&A...595A...1G} so these results will
provide a first check on the \G\ results \citep{2017MNRAS.469..401S}. The
PARSEC observations of objects in the \G\ catalogue will more than double the
temporal baseline allowing the search for unresolved companions at
significantly longer orbits.  The PARSEC results for the objects fainter than
the \G\ limit will remain valuable complementing that mission for science at
the stellar and brown dwarf boundary. Finally, in the case of fainter targets,
the \G\ astrometry of the brighter anonymous field stars will also allow us to
improve reductions using small field astrometry especially in the estimation
of the correction from relative to absolute parallax that remains a constant
floor to what can be achieved with small field astrometry.

\section*{Acknowledgements}

The authors thank the anonymous referee for a thorough review and Amelia Bayo,
Carlos Rodrigo, Jonathan Gagn\'e and Adric Riedel for useful discussions
during the preparation of this manuscript.

This research is based on observations collected at the European Organisation
for Astronomical Research in the Southern Hemisphere, Chile programs
079.A-9203, 081.A-9200, 082.C-0946, 083.C-0446, 085.C-0690, 086.C-0168, and
186.C-0756; in proposal 15B/54 of OPTICON funded under EU FP6 contract number
RII3-CT-001566; through CNTAC in proposal CN2015B--5; the Southern
Astrophysical Research (SOAR) telescope, which is a joint project of the
Minist\'erio da Ci\^encia, Tecnologia, e Inova\c{c}\~ao (MCTI) da
Rep\'{u}blica Federativa do Brasil, the U.S. National Optical Astronomy
Observatory (NOAO), the University of North Carolina at Chapel Hill (UNC), and
Michigan State University (MSU) as part of the proposals SO2009A-008,
SO2011A-009, and SO2011B-006. \newline

RLS was supported by a Henri Chr\'etien International Research Grant
administered by the American Astronomical Society and a Visiting Professorship
with the Leverhulme Trust (VP1-2015-063).  FM/HRAJ/DJP acknowledge support
from the UK's Science and Technology Facilities Council grant number
ST/M001008/1.  FM was supported by an appointment to the NASA Postdoctoral
Program the the Jet Propulsion Laboratory, administered by the Universities
Space Research Association under contract with NASA. AHA/RLS
were supported by the Marie Curie 7th European Community Framework Programme
grant n.236735 {\it Parallaxes of Southern Extremely Cool objects} (PARSEC)
International Incoming Fellowship and grant No.  247593 {\it Interpretation
  and Parameterisation of Extremely Red COOL dwarfs} (IPERCOOL) International
Research Staff Exchange Scheme. RAM acknowledges support from the Chilean
Centro de Excelencia en Astrofisica y Tecnologias Afines (CATA) BASAL PFB/06,
from the Project IC120009 Millennium Institute of Astrophysics of the
Iniciativa Cientifica Milenio del Ministerio de Economia, Fomento y Turismo de
Chile, and from CONICYT/FONDECYT Grant Nr. 117 0854. \newline

This publication makes use of reduction and data products from the 
Cambridge Astronomy Survey Unit (CASU, \url{casu.ast.cam.ac.uk}); 
Centre de Donn\'ees astronomiques de Strasbourg (SIMBAD, \url{cdsweb.u-strasbg.fr}); 
ESA \G\ mission (\url{gea.esac.esa.int/archive/}); 
Panoramic Survey Telescope and Rapid Response System (Pan-STARRS, \url{panstarrs.stsci.edu}); 
Sloan Digital Sky Survey (SDSS, \url{www.sdss.org}); 
SpeX Prism Spectral Libraries (\url{pono.ucsd.edu/~adam/browndwarfs/}); 
Two Micron All Sky Survey (2MASS, \url{www.ipac.caltech.edu/2mass}); 
UKIRT Infrared Deep Sky Survey (UKIDSS, \url{www.ukidss.org}); 
Virtual Observatory SED Analyzer (VOSA, \url{svo2.cab.inta-csic.es/theory/vosa});
Visible and Infrared Survey Telescope for Astronomy surveys (VISTA, \url{horus.roe.ac.uk/vsa}); 
Wide-field Infrared Survey Explorer ($WISE$, \url{wise.ssl.berkeley.edu}).

\bibliographystyle{mnras}
\bibliography{refs}

\begin{thebibliography}{}
\makeatletter
\relax
\def\mn@urlcharsother{\let\do\@makeother \do\$\do\&\do\#\do\^\do\_\do\%\do\~}
\def\mn@doi{\begingroup\mn@urlcharsother \@ifnextchar [ {\mn@doi@}
  {\mn@doi@[]}}
\def\mn@doi@[#1]#2{\def\@tempa{#1}\ifx\@tempa\@empty \href
  {http://dx.doi.org/#2} {doi:#2}\else \href {http://dx.doi.org/#2} {#1}\fi
  \endgroup}
\def\mn@eprint#1#2{\mn@eprint@#1:#2::\@nil}
\def\mn@eprint@arXiv#1{\href {http://arxiv.org/abs/#1} {{\tt arXiv:#1}}}
\def\mn@eprint@dblp#1{\href {http://dblp.uni-trier.de/rec/bibtex/#1.xml}
  {dblp:#1}}
\def\mn@eprint@#1:#2:#3:#4\@nil{\def\@tempa {#1}\def\@tempb {#2}\def\@tempc
  {#3}\ifx \@tempc \@empty \let \@tempc \@tempb \let \@tempb \@tempa \fi \ifx
  \@tempb \@empty \def\@tempb {arXiv}\fi \@ifundefined
  {mn@eprint@\@tempb}{\@tempb:\@tempc}{\expandafter \expandafter \csname
  mn@eprint@\@tempb\endcsname \expandafter{\@tempc}}}

\bibitem[\protect\citeauthoryear{{Ahn} et~al.,}{{Ahn}
  et~al.}{2014}]{2014ApJS..211...17A}
{Ahn} C.~P.,  et~al., 2014, \mn@doi [\apjs] {10.1088/0067-0049/211/2/17}, \href
  {http://cdsads.u-strasbg.fr/abs/2014ApJS..211...17A} {211, 17}

\bibitem[\protect\citeauthoryear{{Allard}, {Homeier}  \& {Freytag}}{{Allard}
  et~al.}{2012}]{2012RSPTA.370.2765A}
{Allard} F.,  {Homeier} D.,   {Freytag} B.,  2012, \mn@doi [Philosophical
  Transactions of the Royal Society of London Series A]
  {10.1098/rsta.2011.0269}, \href
  {http://cdsads.u-strasbg.fr/abs/2012RSPTA.370.2765A} {370, 2765}

\bibitem[\protect\citeauthoryear{{Allers} \& {Liu}}{{Allers} \&
  {Liu}}{2013}]{2013ApJ...772...79A}
{Allers} K.~N.,  {Liu} M.~C.,  2013, \mn@doi [\apj]
  {10.1088/0004-637X/772/2/79}, \href
  {http://adsabs.harvard.edu/abs/2013ApJ...772...79A} {772, 79}

\bibitem[\protect\citeauthoryear{{Andrei} et~al.,}{{Andrei}
  et~al.}{2011}]{2011AJ....141...54A}
{Andrei} A.~H.,  et~al., 2011, \mn@doi [\aj] {10.1088/0004-6256/141/2/54},
  \href {http://cdsads.u-strasbg.fr/abs/2011AJ....141...54A} {141, 54}

\bibitem[\protect\citeauthoryear{{Baade} et~al.,}{{Baade}
  et~al.}{1999}]{1999Msngr..95...15B}
{Baade} D.,  et~al., 1999, The Messenger, \href
  {http://cdsads.u-strasbg.fr/abs/1999Msngr..95...15B} {95, 15}

\bibitem[\protect\citeauthoryear{{Bardalez Gagliuffi} et~al.,}{{Bardalez
  Gagliuffi} et~al.}{2014}]{2014ApJ...794..143B}
{Bardalez Gagliuffi} D.~C.,  et~al., 2014, \mn@doi [\apj]
  {10.1088/0004-637X/794/2/143}, \href
  {http://cdsads.u-strasbg.fr/abs/2014APJ...794..143B} {794, 143}

\bibitem[\protect\citeauthoryear{{Bayo}, {Rodrigo}, {Barrado Y Navascu{\'e}s},
  {Solano}, {Guti{\'e}rrez}, {Morales-Calder{\'o}n}  \& {Allard}}{{Bayo}
  et~al.}{2008}]{2008A&A...492..277B}
{Bayo} A.,  {Rodrigo} C.,  {Barrado Y Navascu{\'e}s} D.,  {Solano} E.,
  {Guti{\'e}rrez} R.,  {Morales-Calder{\'o}n} M.,   {Allard} F.,  2008, \mn@doi
  [\aap] {10.1051/0004-6361:200810395}, \href
  {http://cdsads.u-strasbg.fr/abs/2008A%26A...492..277B} {492, 277}

\bibitem[\protect\citeauthoryear{{Becklin} \& {Zuckerman}}{{Becklin} \&
  {Zuckerman}}{1988}]{1988Natur.336..656B}
{Becklin} E.~E.,  {Zuckerman} B.,  1988, \mn@doi [\nat] {10.1038/336656a0},
  \href {http://cdsads.u-strasbg.fr/abs/1988Natur.336..656B} {336, 656}

\bibitem[\protect\citeauthoryear{{Blake}, {Charbonneau}  \& {White}}{{Blake}
  et~al.}{2010}]{2010ApJ...723..684B}
{Blake} C.~H.,  {Charbonneau} D.,   {White} R.~J.,  2010, \mn@doi [\apj]
  {10.1088/0004-637X/723/1/684}, \href
  {http://cdsads.u-strasbg.fr/abs/2010APJ...723..684B} {723, 684}

\bibitem[\protect\citeauthoryear{{Bouy}, {Brandner}, {Mart{\'{\i}}n},
  {Delfosse}, {Allard}  \& {Basri}}{{Bouy} et~al.}{2003}]{2003AJ....126.1526B}
{Bouy} H.,  {Brandner} W.,  {Mart{\'{\i}}n} E.~L.,  {Delfosse} X.,  {Allard}
  F.,   {Basri} G.,  2003, \mn@doi [\aj] {10.1086/377343}, \href
  {http://cdsads.u-strasbg.fr/abs/2003AJ....126.1526B} {126, 1526}

\bibitem[\protect\citeauthoryear{{Bouy}, {Mart{\'{\i}}n}, {Brandner}  \&
  {Bouvier}}{{Bouy} et~al.}{2005}]{2005AJ....129..511B}
{Bouy} H.,  {Mart{\'{\i}}n} E.~L.,  {Brandner} W.,   {Bouvier} J.,  2005,
  \mn@doi [\aj] {10.1086/426559}, \href
  {http://cdsads.u-strasbg.fr/abs/2005AJ....129..511B} {129, 511}

\bibitem[\protect\citeauthoryear{{Burgasser}}{{Burgasser}}{2004}]{2004ApJ...614L..73B}
{Burgasser} A.~J.,  2004, \mn@doi [\apjl] {10.1086/425418}, \href
  {http://cdsads.u-strasbg.fr/abs/2004APJ...614L..73B} {614, L73}

\bibitem[\protect\citeauthoryear{{Burgasser} et~al.,}{{Burgasser}
  et~al.}{1999}]{1999ApJ...522L..65B}
{Burgasser} A.~J.,  et~al., 1999, \mn@doi [\apjl] {10.1086/312221}, \href
  {http://cdsads.u-strasbg.fr/abs/1999ApJ...522L..65B} {522, L65}

\bibitem[\protect\citeauthoryear{{Burgasser} et~al.,}{{Burgasser}
  et~al.}{2000a}]{2000AJ....120.1100B}
{Burgasser} A.~J.,  et~al., 2000a, \mn@doi [\aj] {10.1086/301475}, \href
  {http://cdsads.u-strasbg.fr/abs/2000AJ....120.1100B} {120, 1100}

\bibitem[\protect\citeauthoryear{{Burgasser} et~al.,}{{Burgasser}
  et~al.}{2000b}]{2000ApJ...531L..57B}
{Burgasser} A.~J.,  et~al., 2000b, \mn@doi [\apjl] {10.1086/312522}, \href
  {http://cdsads.u-strasbg.fr/abs/2000ApJ...531L..57B} {531, L57}

\bibitem[\protect\citeauthoryear{{Burgasser} et~al.,}{{Burgasser}
  et~al.}{2002}]{2002ApJ...564..421B}
{Burgasser} A.~J.,  et~al., 2002, \mn@doi [\apj] {10.1086/324033}, \href
  {http://cdsads.u-strasbg.fr/abs/2002ApJ...564..421B} {564, 421}

\bibitem[\protect\citeauthoryear{{Burgasser}, {McElwain}  \&
  {Kirkpatrick}}{{Burgasser} et~al.}{2003a}]{2003AJ....126.2487B}
{Burgasser} A.~J.,  {McElwain} M.~W.,   {Kirkpatrick} J.~D.,  2003a, \mn@doi
  [\aj] {10.1086/378608}, \href
  {http://cdsads.u-strasbg.fr/abs/2003AJ....126.2487B} {126, 2487}

\bibitem[\protect\citeauthoryear{{Burgasser}, {Kirkpatrick}, {Reid}, {Brown},
  {Miskey}  \& {Gizis}}{{Burgasser} et~al.}{2003b}]{2003ApJ...586..512B}
{Burgasser} A.~J.,  {Kirkpatrick} J.~D.,  {Reid} I.~N.,  {Brown} M.~E.,
  {Miskey} C.~L.,   {Gizis} J.~E.,  2003b, \mn@doi [\apj] {10.1086/346263},
  \href {http://cdsads.u-strasbg.fr/abs/2003ApJ...586..512B} {586, 512}

\bibitem[\protect\citeauthoryear{{Burgasser} et~al.,}{{Burgasser}
  et~al.}{2003c}]{2003ApJ...592.1186B}
{Burgasser} A.~J.,  et~al., 2003c, \mn@doi [\apj] {10.1086/375813}, \href
  {http://cdsads.u-strasbg.fr/abs/2003ApJ...592.1186B} {592, 1186}

\bibitem[\protect\citeauthoryear{{Burgasser}, {Kirkpatrick}, {Cruz}, {Reid},
  {Leggett}, {Liebert}, {Burrows}  \& {Brown}}{{Burgasser}
  et~al.}{2006a}]{2006ApJS..166..585B}
{Burgasser} A.~J.,  {Kirkpatrick} J.~D.,  {Cruz} K.~L.,  {Reid} I.~N.,
  {Leggett} S.~K.,  {Liebert} J.,  {Burrows} A.,   {Brown} M.~E.,  2006a,
  \mn@doi [\apjs] {10.1086/506327}, \href
  {http://cdsads.u-strasbg.fr/abs/2006ApJS..166..585B} {166, 585}

\bibitem[\protect\citeauthoryear{{Burgasser}, {Burrows}  \&
  {Kirkpatrick}}{{Burgasser} et~al.}{2006b}]{2006ApJ...639.1095B}
{Burgasser} A.~J.,  {Burrows} A.,   {Kirkpatrick} J.~D.,  2006b, \mn@doi [\apj]
  {10.1086/499344}, \href {http://cdsads.u-strasbg.fr/abs/2006ApJ...639.1095B}
  {639, 1095}

\bibitem[\protect\citeauthoryear{{Burgasser}, {Looper}, {Kirkpatrick}, {Cruz}
  \& {Swift}}{{Burgasser} et~al.}{2008}]{2008ApJ...674..451B}
{Burgasser} A.~J.,  {Looper} D.~L.,  {Kirkpatrick} J.~D.,  {Cruz} K.~L.,
  {Swift} B.~J.,  2008, \mn@doi [\apj] {10.1086/524726}, \href
  {http://cdsads.u-strasbg.fr/abs/2008ApJ...674..451B} {674, 451}

\bibitem[\protect\citeauthoryear{{Burgasser} et~al.,}{{Burgasser}
  et~al.}{2015}]{2015ApJS..220...18B}
{Burgasser} A.~J.,  et~al., 2015, \mn@doi [\apjs] {10.1088/0067-0049/220/1/18},
  \href {http://adsabs.harvard.edu/abs/2015ApJS..220...18B} {220, 18}

\bibitem[\protect\citeauthoryear{{Burrows}, {Heng}  \& {Nampaisarn}}{{Burrows}
  et~al.}{2011}]{2011ApJ...736...47B}
{Burrows} A.,  {Heng} K.,   {Nampaisarn} T.,  2011, \mn@doi [\apj]
  {10.1088/0004-637X/736/1/47}, \href
  {http://cdsads.u-strasbg.fr/abs/2011ApJ...736...47B} {736, 47}

\bibitem[\protect\citeauthoryear{{Chabrier}, {Baraffe}, {Leconte}, {Gallardo}
  \& {Barman}}{{Chabrier} et~al.}{2009}]{2009AIPC.1094..102C}
{Chabrier} G.,  {Baraffe} I.,  {Leconte} J.,  {Gallardo} J.,   {Barman} T.,
  2009, in {Stempels} E.,  ed.,  American Institute of Physics Conference
  Series Vol. 1094, 15th Cambridge Workshop on Cool Stars, Stellar Systems, and
  the Sun. pp 102--111

\bibitem[\protect\citeauthoryear{{Chambers} et~al.,}{{Chambers}
  et~al.}{2016}]{2016arXiv161205560C}
{Chambers} K.~C.,  et~al., 2016, preprint, \href
  {http://cdsads.u-strasbg.fr/abs/2016arXiv161205560C} {} (\mn@eprint {arXiv}
  {1612.05560})

\bibitem[\protect\citeauthoryear{{Chiu}, {Fan}, {Leggett}, {Golimowski},
  {Zheng}, {Geballe}, {Schneider}  \& {Brinkmann}}{{Chiu}
  et~al.}{2006}]{2006AJ....131.2722C}
{Chiu} K.,  {Fan} X.,  {Leggett} S.~K.,  {Golimowski} D.~A.,  {Zheng} W.,
  {Geballe} T.~R.,  {Schneider} D.~P.,   {Brinkmann} J.,  2006, \mn@doi [\aj]
  {10.1086/501431}, \href {http://cdsads.u-strasbg.fr/abs/2006AJ....131.2722C}
  {131, 2722}

\bibitem[\protect\citeauthoryear{{Cioni} et~al.,}{{Cioni}
  et~al.}{2011}]{2011A&A...527A.116C}
{Cioni} M.-R.~L.,  et~al., 2011, \mn@doi [\aap] {10.1051/0004-6361/201016137},
  \href {http://cdsads.u-strasbg.fr/abs/2011A%26A...527A.116C} {527, A116}

\bibitem[\protect\citeauthoryear{{Cruz}, {Reid}, {Liebert}, {Kirkpatrick}  \&
  {Lowrance}}{{Cruz} et~al.}{2003}]{2003AJ....126.2421C}
{Cruz} K.~L.,  {Reid} I.~N.,  {Liebert} J.,  {Kirkpatrick} J.~D.,   {Lowrance}
  P.~J.,  2003, \mn@doi [\aj] {10.1086/378607}, \href
  {http://cdsads.u-strasbg.fr/abs/2003AJ....126.2421C} {126, 2421}

\bibitem[\protect\citeauthoryear{{Cruz}, {Burgasser}, {Reid}  \&
  {Liebert}}{{Cruz} et~al.}{2004}]{2004ApJ...604L..61C}
{Cruz} K.~L.,  {Burgasser} A.~J.,  {Reid} I.~N.,   {Liebert} J.,  2004, \mn@doi
  [\apjl] {10.1086/383415}, \href
  {http://adsabs.harvard.edu/abs/2004ApJ...604L..61C} {604, L61}

\bibitem[\protect\citeauthoryear{{Cruz} et~al.,}{{Cruz}
  et~al.}{2007}]{2007AJ....133..439C}
{Cruz} K.~L.,  et~al., 2007, \mn@doi [\aj] {10.1086/510132}, \href
  {http://cdsads.u-strasbg.fr/abs/2007AJ....133..439C} {133, 439}

\bibitem[\protect\citeauthoryear{{Cruz}, {Kirkpatrick}  \& {Burgasser}}{{Cruz}
  et~al.}{2009}]{2009AJ....137.3345C}
{Cruz} K.~L.,  {Kirkpatrick} J.~D.,   {Burgasser} A.~J.,  2009, \mn@doi [\aj]
  {10.1088/0004-6256/137/2/3345}, \href
  {http://cdsads.u-strasbg.fr/abs/2009AJ....137.3345C} {137, 3345}

\bibitem[\protect\citeauthoryear{{Dahn} et~al.,}{{Dahn}
  et~al.}{2002}]{2002AJ....124.1170D}
{Dahn} C.~C.,  et~al., 2002, \mn@doi [\aj] {10.1086/341646}, \href
  {http://cdsads.u-strasbg.fr/abs/2002AJ....124.1170D} {124, 1170}

\bibitem[\protect\citeauthoryear{{Delfosse} et~al.,}{{Delfosse}
  et~al.}{1997}]{1997A&A...327L..25D}
{Delfosse} X.,  et~al., 1997, \aap, \href
  {http://cdsads.u-strasbg.fr/abs/1997A%26A...327L..25D} {327, L25}

\bibitem[\protect\citeauthoryear{{Delfosse}, {Tinney}, {Forveille}, {Epchtein},
  {Borsenberger}, {Fouqu{\'e}}, {Kimeswenger}  \& {Tiph{\`e}ne}}{{Delfosse}
  et~al.}{1999}]{1999A&AS..135...41D}
{Delfosse} X.,  {Tinney} C.~G.,  {Forveille} T.,  {Epchtein} N.,
  {Borsenberger} J.,  {Fouqu{\'e}} P.,  {Kimeswenger} S.,   {Tiph{\`e}ne} D.,
  1999, \mn@doi [\aaps] {10.1051/aas:1999158}, \href
  {http://cdsads.u-strasbg.fr/abs/1999A%26AS..135...41D} {135, 41}

\bibitem[\protect\citeauthoryear{{Dieterich}, {Henry}, {Jao}, {Winters},
  {Hosey}, {Riedel}  \& {Subasavage}}{{Dieterich}
  et~al.}{2014}]{2014AJ....147...94D}
{Dieterich} S.~B.,  {Henry} T.~J.,  {Jao} W.-C.,  {Winters} J.~G.,  {Hosey}
  A.~D.,  {Riedel} A.~R.,   {Subasavage} J.~P.,  2014, \mn@doi [\aj]
  {10.1088/0004-6256/147/5/94}, \href
  {http://cdsads.u-strasbg.fr/abs/2014AJ....147...94D} {147, 94}

\bibitem[\protect\citeauthoryear{{Ducourant}, {Teixeira}, {Hambly},
  {Oppenheimer}, {Hawkins}, {Rapaport}, {Modolo}  \& {Lecampion}}{{Ducourant}
  et~al.}{2007}]{2007AA...470..387D}
{Ducourant} C.,  {Teixeira} R.,  {Hambly} N.~C.,  {Oppenheimer} B.~R.,
  {Hawkins} M.~R.~S.,  {Rapaport} M.,  {Modolo} J.,   {Lecampion} J.~F.,  2007,
  \mn@doi [\aap] {10.1051/0004-6361:20066876}, \href
  {http://cdsads.u-strasbg.fr/abs/2007A%26A...470..387D} {470, 387}

\bibitem[\protect\citeauthoryear{{Dupuy} \& {Kraus}}{{Dupuy} \&
  {Kraus}}{2013}]{2013Sci...341.1492D}
{Dupuy} T.~J.,  {Kraus} A.~L.,  2013, \mn@doi [Science]
  {10.1126/science.1241917}, \href
  {http://adsabs.harvard.edu/abs/2013Sci...341.1492D} {341, 1492}

\bibitem[\protect\citeauthoryear{{Dupuy} \& {Liu}}{{Dupuy} \&
  {Liu}}{2012}]{2012ApJS..201...19D}
{Dupuy} T.~J.,  {Liu} M.~C.,  2012, \mn@doi [\apjs]
  {10.1088/0067-0049/201/2/19}, \href
  {http://cdsads.u-strasbg.fr/abs/2012ApJS..201...19D} {201, 19}

\bibitem[\protect\citeauthoryear{{Dupuy} \& {Liu}}{{Dupuy} \&
  {Liu}}{2017}]{2017ApJS..231...15D}
{Dupuy} T.~J.,  {Liu} M.~C.,  2017, \mn@doi [\apjs] {10.3847/1538-4365/aa5e4c},
  \href {http://adsabs.harvard.edu/abs/2017ApJS..231...15D} {231, 15}

\bibitem[\protect\citeauthoryear{{EROS Collaboration} et~al.,}{{EROS
  Collaboration} et~al.}{1999}]{1999A&A...351L...5G}
{EROS Collaboration} et~al., 1999, \aap, \href
  {http://cdsads.u-strasbg.fr/abs/1999A%26A...351L...5G} {351, L5}

\bibitem[\protect\citeauthoryear{{ESA}}{{ESA}}{1997}]{1997HIP...C......0E}
{ESA} 1997, in The Hipparcos and Tycho catalogues. Astrometric and photometric
  star catalogues derived from the ESA Hipparcos Space Astrometry Mission,
  Publisher: Noordwijk, Netherlands: ESA Publications Division, 1997, Series:
  ESA SP Series vol no: 1200, ISBN: 9290923997 (set).

\bibitem[\protect\citeauthoryear{{Faherty}, {Burgasser}, {Cruz}, {Shara},
  {Walter}  \& {Gelino}}{{Faherty} et~al.}{2009}]{2009AJ....137....1F}
{Faherty} J.~K.,  {Burgasser} A.~J.,  {Cruz} K.~L.,  {Shara} M.~M.,  {Walter}
  F.~M.,   {Gelino} C.~R.,  2009, \mn@doi [\aj] {10.1088/0004-6256/137/1/1},
  \href {http://cdsads.u-strasbg.fr/abs/2009AJ....137....1F} {137, 1}

\bibitem[\protect\citeauthoryear{{Faherty} et~al.,}{{Faherty}
  et~al.}{2012}]{2012ApJ...752...56F}
{Faherty} J.~K.,  et~al., 2012, \mn@doi [\apj] {10.1088/0004-637X/752/1/56},
  \href {http://adsabs.harvard.edu/abs/2012ApJ...752...56F} {752, 56}

\bibitem[\protect\citeauthoryear{{Faherty} et~al.,}{{Faherty}
  et~al.}{2016}]{2016ApJS..225...10F}
{Faherty} J.~K.,  et~al., 2016, \mn@doi [\apjs] {10.3847/0067-0049/225/1/10},
  \href {http://cdsads.u-strasbg.fr/abs/2016ApJS..225...10F} {225, 10}

\bibitem[\protect\citeauthoryear{{Fan} et~al.,}{{Fan}
  et~al.}{2000}]{2000AJ....119..928F}
{Fan} X.,  et~al., 2000, \mn@doi [\aj] {10.1086/301224}, \href
  {http://cdsads.u-strasbg.fr/abs/2000AJ....119..928F} {119, 928}

\bibitem[\protect\citeauthoryear{{Folkes}, {Pinfield}, {Kendall}  \&
  {Jones}}{{Folkes} et~al.}{2007}]{2007MNRAS.378..901F}
{Folkes} S.~L.,  {Pinfield} D.~J.,  {Kendall} T.~R.,   {Jones} H.~R.~A.,  2007,
  \mn@doi [\mnras] {10.1111/j.1365-2966.2007.11789.x}, \href
  {http://cdsads.u-strasbg.fr/abs/2007MNRAS.378..901F} {378, 901}

\bibitem[\protect\citeauthoryear{{Francis} \& {Anderson}}{{Francis} \&
  {Anderson}}{2009}]{2009NewA...14..615F}
{Francis} C.,  {Anderson} E.,  2009, \mn@doi [\na]
  {10.1016/j.newast.2009.03.004}, \href
  {http://cdsads.u-strasbg.fr/abs/2009NewA...14..615F} {14, 615}

\bibitem[\protect\citeauthoryear{{Gagn{\'e}}, {Lafreni{\`e}re}, {Doyon}, {Malo}
   \& {Artigau}}{{Gagn{\'e}} et~al.}{2014}]{2014ApJ...783..121G}
{Gagn{\'e}} J.,  {Lafreni{\`e}re} D.,  {Doyon} R.,  {Malo} L.,   {Artigau}
  {\'E}.,  2014, \mn@doi [\apj] {10.1088/0004-637X/783/2/121}, \href
  {http://cdsads.u-strasbg.fr/abs/2014APJ...783..121G} {783, 121}

\bibitem[\protect\citeauthoryear{{Gagn{\'e}} et~al.,}{{Gagn{\'e}}
  et~al.}{2015a}]{2015ApJS..219...33G}
{Gagn{\'e}} J.,  et~al., 2015a, \mn@doi [\apjs] {10.1088/0067-0049/219/2/33},
  \href {http://cdsads.u-strasbg.fr/abs/2015ApJS..219...33G} {219, 33}

\bibitem[\protect\citeauthoryear{{Gagn{\'e}}, {Lafreni{\`e}re}, {Doyon}, {Malo}
   \& {Artigau}}{{Gagn{\'e}} et~al.}{2015b}]{2015ApJ...798...73G}
{Gagn{\'e}} J.,  {Lafreni{\`e}re} D.,  {Doyon} R.,  {Malo} L.,   {Artigau}
  {\'E}.,  2015b, \mn@doi [\apj] {10.1088/0004-637X/798/2/73}, \href
  {http://cdsads.u-strasbg.fr/abs/2015ApJ...798...73G} {798, 73}

\bibitem[\protect\citeauthoryear{{Gagn{\'e}} et~al.,}{{Gagn{\'e}}
  et~al.}{2018}]{2018ApJ...856...23G}
{Gagn{\'e}} J.,  et~al., 2018, \mn@doi [\apj] {10.3847/1538-4357/aaae09}, \href
  {http://adsabs.harvard.edu/abs/2018ApJ...856...23G} {856, 23}

\bibitem[\protect\citeauthoryear{{Gaia Collaboration} et~al.,}{{Gaia
  Collaboration} et~al.}{2016a}]{2016A&A...595A...1G}
{Gaia Collaboration} et~al., 2016a, \mn@doi [\aap]
  {10.1051/0004-6361/201629272}, \href
  {http://adsabs.harvard.edu/abs/2016A%26A...595A...1G} {595, A1}

\bibitem[\protect\citeauthoryear{{Gaia Collaboration} et~al.,}{{Gaia
  Collaboration} et~al.}{2016b}]{2016A&A...595A...2G}
{Gaia Collaboration} et~al., 2016b, \mn@doi [\aap]
  {10.1051/0004-6361/201629512}, \href
  {http://cdsads.u-strasbg.fr/abs/2016A%26A...595A...2G} {595, A2}

\bibitem[\protect\citeauthoryear{{G{\'a}lvez-Ortiz} et~al.,}{{G{\'a}lvez-Ortiz}
  et~al.}{2010}]{2010MNRAS.409..552G}
{G{\'a}lvez-Ortiz} M.~C.,  et~al., 2010, \mn@doi [\mnras]
  {10.1111/j.1365-2966.2010.17361.x}, \href
  {http://cdsads.u-strasbg.fr/abs/2010MNRAS.409..552G} {409, 552}

\bibitem[\protect\citeauthoryear{{Geballe} et~al.,}{{Geballe}
  et~al.}{2002}]{2002ApJ...564..466G}
{Geballe} T.~R.,  et~al., 2002, \mn@doi [\apj] {10.1086/324078}, \href
  {http://adsabs.harvard.edu/abs/2002ApJ...564..466G} {564, 466}

\bibitem[\protect\citeauthoryear{{Gelino} \& {Burgasser}}{{Gelino} \&
  {Burgasser}}{2010}]{2010AJ....140..110G}
{Gelino} C.~R.,  {Burgasser} A.~J.,  2010, \mn@doi [\aj]
  {10.1088/0004-6256/140/1/110}, \href
  {http://cdsads.u-strasbg.fr/abs/2010AJ....140..110G} {140, 110}

\bibitem[\protect\citeauthoryear{{Gizis}}{{Gizis}}{2002}]{2002ApJ...575..484G}
{Gizis} J.~E.,  2002, \mn@doi [\apj] {10.1086/341259}, \href
  {http://cdsads.u-strasbg.fr/abs/2002ApJ...575..484G} {575, 484}

\bibitem[\protect\citeauthoryear{{Gizis}, {Kirkpatrick}  \& {Wilson}}{{Gizis}
  et~al.}{2001}]{2001AJ....121.2185G}
{Gizis} J.~E.,  {Kirkpatrick} J.~D.,   {Wilson} J.~C.,  2001, \mn@doi [\aj]
  {10.1086/319937}, \href {http://cdsads.u-strasbg.fr/abs/2001AJ....121.2185G}
  {121, 2185}

\bibitem[\protect\citeauthoryear{{Golimowski} et~al.,}{{Golimowski}
  et~al.}{2004}]{2004AJ....128.1733G}
{Golimowski} D.~A.,  et~al., 2004, \mn@doi [\aj] {10.1086/423911}, \href
  {http://cdsads.u-strasbg.fr/cgi-bin/nph-bib_query?bibcode=2004AJ....128.1733G&db_key=AST}
  {128, 1733}

\bibitem[\protect\citeauthoryear{{Guenther} \& {Wuchterl}}{{Guenther} \&
  {Wuchterl}}{2003}]{2003A&A...401..677G}
{Guenther} E.~W.,  {Wuchterl} G.,  2003, \mn@doi [\aap]
  {10.1051/0004-6361:20030149}, \href
  {http://cdsads.u-strasbg.fr/abs/2003A%26A...401..677G} {401, 677}

\bibitem[\protect\citeauthoryear{{Hawley} et~al.,}{{Hawley}
  et~al.}{2002}]{2002AJ....123.3409H}
{Hawley} S.~L.,  et~al., 2002, \mn@doi [\aj] {10.1086/340697}, \href
  {http://adsabs.harvard.edu/abs/2002AJ....123.3409H} {123, 3409}

\bibitem[\protect\citeauthoryear{{Henry}, {Jao}, {Subasavage}, {Beaulieu},
  {Ianna}, {Costa}  \& {M{\'e}ndez}}{{Henry}
  et~al.}{2006}]{2006AJ....132.2360H}
{Henry} T.~J.,  {Jao} W.-C.,  {Subasavage} J.~P.,  {Beaulieu} T.~D.,  {Ianna}
  P.~A.,  {Costa} E.,   {M{\'e}ndez} R.~A.,  2006, \mn@doi [\aj]
  {10.1086/508233}, \href {http://cdsads.u-strasbg.fr/abs/2006AJ....132.2360H}
  {132, 2360}

\bibitem[\protect\citeauthoryear{{Jameson}, {Casewell}, {Bannister}, {Lodieu},
  {Keresztes}, {Dobbie}  \& {Hodgkin}}{{Jameson}
  et~al.}{2008}]{2008MNRAS.384.1399J}
{Jameson} R.~F.,  {Casewell} S.~L.,  {Bannister} N.~P.,  {Lodieu} N.,
  {Keresztes} K.,  {Dobbie} P.~D.,   {Hodgkin} S.~T.,  2008, \mn@doi [\mnras]
  {10.1111/j.1365-2966.2007.12637.x}, \href
  {http://cdsads.u-strasbg.fr/abs/2008MNRAS.384.1399J} {384, 1399}

\bibitem[\protect\citeauthoryear{{Johnson} \& {Soderblom}}{{Johnson} \&
  {Soderblom}}{1987}]{1987AJ.....93..864J}
{Johnson} D.~R.~H.,  {Soderblom} D.~R.,  1987, \mn@doi [\aj] {10.1086/114370},
  \href {http://cdsads.u-strasbg.fr/abs/1987AJ.....93..864J} {93, 864}

\bibitem[\protect\citeauthoryear{{Kendall}, {Mauron}, {Azzopardi}  \&
  {Gigoyan}}{{Kendall} et~al.}{2003}]{2003A&A...403..929K}
{Kendall} T.~R.,  {Mauron} N.,  {Azzopardi} M.,   {Gigoyan} K.,  2003, \mn@doi
  [\aap] {10.1051/0004-6361:20030218}, \href
  {http://cdsads.u-strasbg.fr/abs/2003A%26A...403..929K} {403, 929}

\bibitem[\protect\citeauthoryear{{Kendall}, {Delfosse}, {Mart{\'{\i}}n}  \&
  {Forveille}}{{Kendall} et~al.}{2004}]{2004A&A...416L..17K}
{Kendall} T.~R.,  {Delfosse} X.,  {Mart{\'{\i}}n} E.~L.,   {Forveille} T.,
  2004, \mn@doi [\aap] {10.1051/0004-6361:20040046}, \href
  {http://cdsads.u-strasbg.fr/abs/2004A%26A...416L..17K} {416, L17}

\bibitem[\protect\citeauthoryear{{Kendall} et~al.,}{{Kendall}
  et~al.}{2007}]{2007A&A...466.1059K}
{Kendall} T.~R.,  et~al., 2007, \mn@doi [\aap] {10.1051/0004-6361:20066403},
  \href {http://cdsads.u-strasbg.fr/abs/2007A%26A...466.1059K} {466, 1059}

\bibitem[\protect\citeauthoryear{{Kirkpatrick} et~al.,}{{Kirkpatrick}
  et~al.}{1999}]{1999ApJ...519..802K}
{Kirkpatrick} J.~D.,  et~al., 1999, \mn@doi [\apj] {10.1086/307414}, \href
  {http://cdsads.u-strasbg.fr/abs/1999ApJ...519..802K} {519, 802}

\bibitem[\protect\citeauthoryear{{Kirkpatrick} et~al.,}{{Kirkpatrick}
  et~al.}{2000}]{2000AJ....120..447K}
{Kirkpatrick} J.~D.,  et~al., 2000, \mn@doi [\aj] {10.1086/301427}, \href
  {http://cdsads.u-strasbg.fr/abs/2000AJ....120..447K} {120, 447}

\bibitem[\protect\citeauthoryear{{Kirkpatrick}, {Dahn}, {Monet}, {Reid},
  {Gizis}, {Liebert}  \& {Burgasser}}{{Kirkpatrick}
  et~al.}{2001}]{2001AJ....121.3235K}
{Kirkpatrick} J.~D.,  {Dahn} C.~C.,  {Monet} D.~G.,  {Reid} I.~N.,  {Gizis}
  J.~E.,  {Liebert} J.,   {Burgasser} A.~J.,  2001, \mn@doi [\aj]
  {10.1086/321085}, \href {http://cdsads.u-strasbg.fr/abs/2001AJ....121.3235K}
  {121, 3235}

\bibitem[\protect\citeauthoryear{{Kirkpatrick} et~al.,}{{Kirkpatrick}
  et~al.}{2008}]{2008ApJ...689.1295K}
{Kirkpatrick} J.~D.,  et~al., 2008, \mn@doi [\apj] {10.1086/592768}, \href
  {http://cdsads.u-strasbg.fr/abs/2008ApJ...689.1295K} {689, 1295}

\bibitem[\protect\citeauthoryear{{Knapp} et~al.,}{{Knapp}
  et~al.}{2004}]{2004AJ....127.3553K}
{Knapp} G.~R.,  et~al., 2004, \mn@doi [\aj] {10.1086/420707}, \href
  {http://adsabs.harvard.edu/abs/2004AJ....127.3553K} {127, 3553}

\bibitem[\protect\citeauthoryear{{LSST Science Collaboration} et~al.,}{{LSST
  Science Collaboration} et~al.}{2017}]{2017arXiv170804058L}
{LSST Science Collaboration} et~al., 2017, preprint, \href
  {http://cdsads.u-strasbg.fr/abs/2017arXiv170804058L} {} (\mn@eprint {arXiv}
  {1708.04058})

\bibitem[\protect\citeauthoryear{{Latham}, {Stefanik}, {Mazeh}, {Mayor}  \&
  {Burki}}{{Latham} et~al.}{1989}]{1989Natur.339...38L}
{Latham} D.~W.,  {Stefanik} R.~P.,  {Mazeh} T.,  {Mayor} M.,   {Burki} G.,
  1989, \mn@doi [\nat] {10.1038/339038a0}, \href
  {http://adsabs.harvard.edu/abs/1989Natur.339...38L} {339, 38}

\bibitem[\protect\citeauthoryear{{Laureijs}, {Duvet}, {Escudero Sanz},
  {Gondoin}, {Lumb}, {Oosterbroek}  \& {Saavedra Criado}}{{Laureijs}
  et~al.}{2010}]{2010SPIE.7731E..1HL}
{Laureijs} R.~J.,  {Duvet} L.,  {Escudero Sanz} I.,  {Gondoin} P.,  {Lumb}
  D.~H.,  {Oosterbroek} T.,   {Saavedra Criado} G.,  2010, in Space Telescopes
  and Instrumentation 2010: Optical, Infrared, and Millimeter Wave. p. 77311H,
  \mn@doi{10.1117/12.857123}

\bibitem[\protect\citeauthoryear{{Lawrence} et~al.,}{{Lawrence}
  et~al.}{2007}]{2007MNRAS.379.1599L}
{Lawrence} A.,  et~al., 2007, \mn@doi [\mnras]
  {10.1111/j.1365-2966.2007.12040.x}, \href
  {http://adsabs.harvard.edu/abs/2007MNRAS.379.1599L} {379, 1599}

\bibitem[\protect\citeauthoryear{{Leggett} et~al.,}{{Leggett}
  et~al.}{2000}]{2000ApJ...536L..35L}
{Leggett} S.~K.,  et~al., 2000, \mn@doi [\apjl] {10.1086/312728}, \href
  {http://cdsads.u-strasbg.fr/abs/2000APJ...536L..35L} {536, L35}

\bibitem[\protect\citeauthoryear{{L{\'e}pine}, {Hilton}, {Mann}, {Wilde},
  {Rojas-Ayala}, {Cruz}  \& {Gaidos}}{{L{\'e}pine}
  et~al.}{2013}]{2013AJ....145..102L}
{L{\'e}pine} S.,  {Hilton} E.~J.,  {Mann} A.~W.,  {Wilde} M.,  {Rojas-Ayala}
  B.,  {Cruz} K.~L.,   {Gaidos} E.,  2013, \mn@doi [\aj]
  {10.1088/0004-6256/145/4/102}, \href
  {http://cdsads.u-strasbg.fr/abs/2013AJ....145..102L} {145, 102}

\bibitem[\protect\citeauthoryear{{Liebert}, {Kirkpatrick}, {Cruz}, {Reid},
  {Burgasser}, {Tinney}  \& {Gizis}}{{Liebert}
  et~al.}{2003}]{2003AJ....125..343L}
{Liebert} J.,  {Kirkpatrick} J.~D.,  {Cruz} K.~L.,  {Reid} I.~N.,  {Burgasser}
  A.,  {Tinney} C.~G.,   {Gizis} J.~E.,  2003, \mn@doi [\aj] {10.1086/345514},
  \href {http://cdsads.u-strasbg.fr/abs/2003AJ....125..343L} {125, 343}

\bibitem[\protect\citeauthoryear{{Liu}, {Dupuy}  \& {Allers}}{{Liu}
  et~al.}{2016}]{2016ApJ...833...96L}
{Liu} M.~C.,  {Dupuy} T.~J.,   {Allers} K.~N.,  2016, \mn@doi [\apj]
  {10.3847/1538-4357/833/1/96}, \href
  {http://cdsads.u-strasbg.fr/abs/2016ApJ...833...96L} {833, 96}

\bibitem[\protect\citeauthoryear{{Lodieu}, {Scholz}  \& {McCaughrean}}{{Lodieu}
  et~al.}{2002}]{2002A&A...389L..20L}
{Lodieu} N.,  {Scholz} R.-D.,   {McCaughrean} M.~J.,  2002, \mn@doi [\aap]
  {10.1051/0004-6361:20020698}, \href
  {http://cdsads.u-strasbg.fr/abs/2002A%26A...389L..20L} {389, L20}

\bibitem[\protect\citeauthoryear{{Lodieu}, {Scholz}, {McCaughrean}, {Ibata},
  {Irwin}  \& {Zinnecker}}{{Lodieu} et~al.}{2005}]{2005A&A...440.1061L}
{Lodieu} N.,  {Scholz} R.-D.,  {McCaughrean} M.~J.,  {Ibata} R.,  {Irwin} M.,
  {Zinnecker} H.,  2005, \mn@doi [\aap] {10.1051/0004-6361:20042456}, \href
  {http://cdsads.u-strasbg.fr/abs/2005A%26A...440.1061L} {440, 1061}

\bibitem[\protect\citeauthoryear{{Looper}, {Kirkpatrick}  \&
  {Burgasser}}{{Looper} et~al.}{2007}]{2007AJ....134.1162L}
{Looper} D.~L.,  {Kirkpatrick} J.~D.,   {Burgasser} A.~J.,  2007, \mn@doi [\aj]
  {10.1086/520645}, \href {http://cdsads.u-strasbg.fr/abs/2007AJ....134.1162L}
  {134, 1162}

\bibitem[\protect\citeauthoryear{{Looper}, {Gelino}, {Burgasser}  \&
  {Kirkpatrick}}{{Looper} et~al.}{2008}]{2008ApJ...685.1183L}
{Looper} D.~L.,  {Gelino} C.~R.,  {Burgasser} A.~J.,   {Kirkpatrick} J.~D.,
  2008, \mn@doi [\apj] {10.1086/590382}, \href
  {http://cdsads.u-strasbg.fr/abs/2008ApJ...685.1183L} {685, 1183}

\bibitem[\protect\citeauthoryear{{Mace} et~al.,}{{Mace}
  et~al.}{2013}]{2013ApJS..205....6M}
{Mace} G.~N.,  et~al., 2013, \mn@doi [\apjs] {10.1088/0067-0049/205/1/6}, \href
  {http://cdsads.u-strasbg.fr/abs/2013ApJS..205....6M} {205, 6}

\bibitem[\protect\citeauthoryear{{Ma{\'{\i}}z Apell{\'a}niz}}{{Ma{\'{\i}}z
  Apell{\'a}niz}}{2017}]{2017A&A...608L...8M}
{Ma{\'{\i}}z Apell{\'a}niz} J.,  2017, \mn@doi [\aap]
  {10.1051/0004-6361/201732167}, \href
  {http://cdsads.u-strasbg.fr/abs/2017A%26A...608L...8M} {608, L8}

\bibitem[\protect\citeauthoryear{{Malo}, {Doyon}, {Lafreni{\`e}re}, {Artigau},
  {Gagn{\'e}}, {Baron}  \& {Riedel}}{{Malo} et~al.}{2013}]{2013ApJ...762...88M}
{Malo} L.,  {Doyon} R.,  {Lafreni{\`e}re} D.,  {Artigau} {\'E}.,  {Gagn{\'e}}
  J.,  {Baron} F.,   {Riedel} A.,  2013, \mn@doi [\apj]
  {10.1088/0004-637X/762/2/88}, \href
  {http://adsabs.harvard.edu/abs/2013ApJ...762...88M} {762, 88}

\bibitem[\protect\citeauthoryear{{Manjavacas} et~al.,}{{Manjavacas}
  et~al.}{2016}]{2016MNRAS.455.1341M}
{Manjavacas} E.,  et~al., 2016, \mn@doi [\mnras] {10.1093/mnras/stv2048}, \href
  {http://cdsads.u-strasbg.fr/abs/2016MNRAS.455.1341M} {455, 1341}

\bibitem[\protect\citeauthoryear{{Marocco} et~al.,}{{Marocco}
  et~al.}{2013}]{2013AJ....146..161M}
{Marocco} F.,  et~al., 2013, \mn@doi [\aj] {10.1088/0004-6256/146/6/161}, \href
  {http://cdsads.u-strasbg.fr/abs/2013AJ....146..161M} {146, 161}

\bibitem[\protect\citeauthoryear{{Mart{\'{\i}}n}, {Delfosse}, {Basri},
  {Goldman}, {Forveille}  \& {Zapatero Osorio}}{{Mart{\'{\i}}n}
  et~al.}{1999}]{1999AJ....118.2466M}
{Mart{\'{\i}}n} E.~L.,  {Delfosse} X.,  {Basri} G.,  {Goldman} B.,  {Forveille}
  T.,   {Zapatero Osorio} M.~R.,  1999, \mn@doi [\aj] {10.1086/301107}, \href
  {http://cdsads.u-strasbg.fr/abs/1999AJ....118.2466M} {118, 2466}

\bibitem[\protect\citeauthoryear{{Mart{\'{\i}}n} et~al.,}{{Mart{\'{\i}}n}
  et~al.}{2010}]{2010A&A...517A..53M}
{Mart{\'{\i}}n} E.~L.,  et~al., 2010, \mn@doi [\aap]
  {10.1051/0004-6361/201014202}, \href
  {http://cdsads.u-strasbg.fr/abs/2010A%26A...517A..53M} {517, A53}

\bibitem[\protect\citeauthoryear{{Mason}, {Wycoff}, {Hartkopf}, {Douglass}  \&
  {Worley}}{{Mason} et~al.}{2001}]{2001AJ....122.3466M}
{Mason} B.~D.,  {Wycoff} G.~L.,  {Hartkopf} W.~I.,  {Douglass} G.~G.,
  {Worley} C.~E.,  2001, \mn@doi [\aj] {10.1086/323920}, \href
  {http://cdsads.u-strasbg.fr/abs/2001AJ....122.3466M} {122, 3466}

\bibitem[\protect\citeauthoryear{{McElwain} \& {Burgasser}}{{McElwain} \&
  {Burgasser}}{2006}]{2006AJ....132.2074M}
{McElwain} M.~W.,  {Burgasser} A.~J.,  2006, \mn@doi [\aj] {10.1086/508199},
  \href {http://cdsads.u-strasbg.fr/abs/2006AJ....132.2074M} {132, 2074}

\bibitem[\protect\citeauthoryear{{McMahon}, {Banerji}, {Gonzalez}, {Koposov},
  {Bejar}, {Lodieu}, {Rebolo}  \& {VHS Collaboration}}{{McMahon}
  et~al.}{2013}]{2013Msngr.154...35M}
{McMahon} R.~G.,  {Banerji} M.,  {Gonzalez} E.,  {Koposov} S.~E.,  {Bejar}
  V.~J.,  {Lodieu} N.,  {Rebolo} R.,   {VHS Collaboration} 2013, The Messenger,
  \href {http://cdsads.u-strasbg.fr/abs/2013Msngr.154...35M} {154, 35}

\bibitem[\protect\citeauthoryear{{Mendez} \& {van Altena}}{{Mendez} \& {van
  Altena}}{1996}]{men96}
{Mendez} R.~A.,  {van Altena} W.~F.,  1996, \aj, 112, 655

\bibitem[\protect\citeauthoryear{{Miles-P{\'a}ez}, {Metchev}, {Heinze}  \&
  {Apai}}{{Miles-P{\'a}ez} et~al.}{2017}]{2017ApJ...840...83M}
{Miles-P{\'a}ez} P.~A.,  {Metchev} S.~A.,  {Heinze} A.,   {Apai} D.,  2017,
  \mn@doi [\apj] {10.3847/1538-4357/aa6f11}, \href
  {http://cdsads.u-strasbg.fr/abs/2017ApJ...840...83M} {840, 83}

\bibitem[\protect\citeauthoryear{{Minniti} et~al.,}{{Minniti}
  et~al.}{2010}]{2010NewA...15..433M}
{Minniti} D.,  et~al., 2010, \mn@doi [\na] {10.1016/j.newast.2009.12.002},
  \href {http://cdsads.u-strasbg.fr/abs/2010NewA...15..433M} {15, 433}

\bibitem[\protect\citeauthoryear{{Monet}, {Dahn}, {Vrba}, {Harris}, {Pier},
  {Luginbuhl}  \& {Ables}}{{Monet} et~al.}{1992}]{1992AJ....103..638M}
{Monet} D.~G.,  {Dahn} C.~C.,  {Vrba} F.~J.,  {Harris} H.~C.,  {Pier} J.~R.,
  {Luginbuhl} C.~B.,   {Ables} H.~D.,  1992, \mn@doi [\aj] {10.1086/116091},
  \href {http://cdsads.u-strasbg.fr/abs/1992AJ....103..638M} {103, 638}

\bibitem[\protect\citeauthoryear{{Paudel}, {Gizis}, {Mullan}, {Schmidt},
  {Burgasser}, {Williams}  \& {Berger}}{{Paudel}
  et~al.}{2018}]{2018ApJ...861...76P}
{Paudel} R.~R.,  {Gizis} J.~E.,  {Mullan} D.~J.,  {Schmidt} S.~J.,  {Burgasser}
  A.~J.,  {Williams} P.~K.~G.,   {Berger} E.,  2018, \mn@doi [\apj]
  {10.3847/1538-4357/aac8e0}, \href
  {http://cdsads.u-strasbg.fr/abs/2018ApJ...861...76P} {861, 76}

\bibitem[\protect\citeauthoryear{{Reid}, {Kirkpatrick}, {Gizis}, {Dahn},
  {Monet}, {Williams}, {Liebert}  \& {Burgasser}}{{Reid}
  et~al.}{2000}]{2000AJ....119..369R}
{Reid} I.~N.,  {Kirkpatrick} J.~D.,  {Gizis} J.~E.,  {Dahn} C.~C.,  {Monet}
  D.~G.,  {Williams} R.~J.,  {Liebert} J.,   {Burgasser} A.~J.,  2000, \mn@doi
  [\aj] {10.1086/301177}, \href
  {http://cdsads.u-strasbg.fr/abs/2000AJ....119..369R} {119, 369}

\bibitem[\protect\citeauthoryear{{Reid}, {Lewitus}, {Allen}, {Cruz}  \&
  {Burgasser}}{{Reid} et~al.}{2006a}]{2006AJ....132..891R}
{Reid} I.~N.,  {Lewitus} E.,  {Allen} P.~R.,  {Cruz} K.~L.,   {Burgasser}
  A.~J.,  2006a, \mn@doi [\aj] {10.1086/505626}, \href
  {http://cdsads.u-strasbg.fr/abs/2006AJ....132..891R} {132, 891}

\bibitem[\protect\citeauthoryear{{Reid}, {Lewitus}, {Burgasser}  \&
  {Cruz}}{{Reid} et~al.}{2006b}]{2006ApJ...639.1114R}
{Reid} I.~N.,  {Lewitus} E.,  {Burgasser} A.~J.,   {Cruz} K.~L.,  2006b,
  \mn@doi [\apj] {10.1086/499484}, \href
  {http://cdsads.u-strasbg.fr/abs/2006ApJ...639.1114R} {639, 1114}

\bibitem[\protect\citeauthoryear{{Reid}, {Cruz}, {Kirkpatrick}, {Allen},
  {Mungall}, {Liebert}, {Lowrance}  \& {Sweet}}{{Reid}
  et~al.}{2008}]{2008AJ....136.1290R}
{Reid} I.~N.,  {Cruz} K.~L.,  {Kirkpatrick} J.~D.,  {Allen} P.~R.,  {Mungall}
  F.,  {Liebert} J.,  {Lowrance} P.,   {Sweet} A.,  2008, \mn@doi [\aj]
  {10.1088/0004-6256/136/3/1290}, \href
  {http://cdsads.u-strasbg.fr/abs/2008AJ....136.1290R} {136, 1290}

\bibitem[\protect\citeauthoryear{{Reiners} \& {Basri}}{{Reiners} \&
  {Basri}}{2009}]{2009ApJ...705.1416R}
{Reiners} A.,  {Basri} G.,  2009, \mn@doi [\apj]
  {10.1088/0004-637X/705/2/1416}, \href
  {http://cdsads.u-strasbg.fr/abs/2009APJ...705.1416R} {705, 1416}

\bibitem[\protect\citeauthoryear{{Riedel}, {Blunt}, {Lambrides}, {Rice}, {Cruz}
   \& {Faherty}}{{Riedel} et~al.}{2017}]{2017AJ....153...95R}
{Riedel} A.~R.,  {Blunt} S.~C.,  {Lambrides} E.~L.,  {Rice} E.~L.,  {Cruz}
  K.~L.,   {Faherty} J.~K.,  2017, \mn@doi [\aj] {10.3847/1538-3881/153/3/95},
  \href {http://adsabs.harvard.edu/abs/2017AJ....153...95R} {153, 95}

\bibitem[\protect\citeauthoryear{{Sahlmann}, {Lazorenko}, {S{\'e}gransan},
  {Mart{\'{\i}}n}, {Mayor}, {Queloz}  \& {Udry}}{{Sahlmann}
  et~al.}{2014}]{2014A&A...565A..20S}
{Sahlmann} J.,  {Lazorenko} P.~F.,  {S{\'e}gransan} D.,  {Mart{\'{\i}}n} E.~L.,
   {Mayor} M.,  {Queloz} D.,   {Udry} S.,  2014, \mn@doi [\aap]
  {10.1051/0004-6361/201323208}, \href
  {http://cdsads.u-strasbg.fr/abs/2014A%26A...565A..20S} {565, A20}

\bibitem[\protect\citeauthoryear{{Sarro}, {Berihuete}, {Carri{\'o}n},
  {Barrado}, {Cruz}  \& {Isasi}}{{Sarro} et~al.}{2013}]{2013A&A...550A..44S}
{Sarro} L.~M.,  {Berihuete} A.,  {Carri{\'o}n} C.,  {Barrado} D.,  {Cruz} P.,
  {Isasi} Y.,  2013, \mn@doi [\aap] {10.1051/0004-6361/201219867}, \href
  {http://adsabs.harvard.edu/abs/2013A%26A...550A..44S} {550, A44}

\bibitem[\protect\citeauthoryear{{Schmidt}, {West}, {Hawley}  \&
  {Pineda}}{{Schmidt} et~al.}{2010}]{2010AJ....139.1808S}
{Schmidt} S.~J.,  {West} A.~A.,  {Hawley} S.~L.,   {Pineda} J.~S.,  2010,
  \mn@doi [\aj] {10.1088/0004-6256/139/5/1808}, \href
  {http://cdsads.u-strasbg.fr/abs/2010AJ....139.1808S} {139, 1808}

\bibitem[\protect\citeauthoryear{{Schneider} et~al.,}{{Schneider}
  et~al.}{2002}]{2002AJ....123..458S}
{Schneider} D.~P.,  et~al., 2002, \mn@doi [\aj] {10.1086/338095}, \href
  {http://cdsads.u-strasbg.fr/abs/2002AJ....123..458S} {123, 458}

\bibitem[\protect\citeauthoryear{{Schneider}, {Cushing}, {Kirkpatrick}, {Mace},
  {Gelino}, {Faherty}, {Fajardo-Acosta}  \& {Sheppard}}{{Schneider}
  et~al.}{2014}]{2014AJ....147...34S}
{Schneider} A.~C.,  {Cushing} M.~C.,  {Kirkpatrick} J.~D.,  {Mace} G.~N.,
  {Gelino} C.~R.,  {Faherty} J.~K.,  {Fajardo-Acosta} S.,   {Sheppard} S.~S.,
  2014, \mn@doi [\aj] {10.1088/0004-6256/147/2/34}, \href
  {http://cdsads.u-strasbg.fr/abs/2014AJ....147...34S} {147, 34}

\bibitem[\protect\citeauthoryear{{Scholz} \& {Meusinger}}{{Scholz} \&
  {Meusinger}}{2002}]{2002MNRAS.336L..49S}
{Scholz} R.-D.,  {Meusinger} H.,  2002, \mn@doi [\mnras]
  {10.1046/j.1365-8711.2002.05998.x}, \href
  {http://cdsads.u-strasbg.fr/abs/2002MNRAS.336L..49S} {336, L49}

\bibitem[\protect\citeauthoryear{{Scholz}, {McCaughrean}, {Lodieu}  \&
  {Kuhlbrodt}}{{Scholz} et~al.}{2003}]{2003A&A...398L..29S}
{Scholz} R.-D.,  {McCaughrean} M.~J.,  {Lodieu} N.,   {Kuhlbrodt} B.,  2003,
  \mn@doi [\aap] {10.1051/0004-6361:20021847}, \href
  {http://cdsads.u-strasbg.fr/abs/2003A%26A...398L..29S} {398, L29}

\bibitem[\protect\citeauthoryear{{Scholz}, {Lehmann}, {Matute}  \&
  {Zinnecker}}{{Scholz} et~al.}{2004}]{2004A&A...425..519S}
{Scholz} R.-D.,  {Lehmann} I.,  {Matute} I.,   {Zinnecker} H.,  2004, \mn@doi
  [\aap] {10.1051/0004-6361:20041059}, \href
  {http://cdsads.u-strasbg.fr/abs/2004A%26A...425..519S} {425, 519}

\bibitem[\protect\citeauthoryear{{Sch{\"o}nrich}, {Binney}  \&
  {Dehnen}}{{Sch{\"o}nrich} et~al.}{2010}]{2010MNRAS.403.1829S}
{Sch{\"o}nrich} R.,  {Binney} J.,   {Dehnen} W.,  2010, \mn@doi [\mnras]
  {10.1111/j.1365-2966.2010.16253.x}, \href
  {http://cdsads.u-strasbg.fr/abs/2010MNRAS.403.1829S} {403, 1829}

\bibitem[\protect\citeauthoryear{{Seifahrt}, {Mugrauer}, {Wiese},
  {Neuh{\"a}user}  \& {Guenther}}{{Seifahrt}
  et~al.}{2005a}]{2005AN....326..974S}
{Seifahrt} A.,  {Mugrauer} M.,  {Wiese} M.,  {Neuh{\"a}user} R.,   {Guenther}
  E.~W.,  2005a, \mn@doi [Astronomische Nachrichten] {10.1002/asna.200510456},
  \href {http://cdsads.u-strasbg.fr/abs/2005AN....326..974S} {326, 974}

\bibitem[\protect\citeauthoryear{{Seifahrt}, {Guenther}  \&
  {Neuh{\"a}user}}{{Seifahrt} et~al.}{2005b}]{2005A&A...440..967S}
{Seifahrt} A.,  {Guenther} E.,   {Neuh{\"a}user} R.,  2005b, \mn@doi [\aap]
  {10.1051/0004-6361:20041902}, \href
  {http://cdsads.u-strasbg.fr/abs/2005A%26A...440..967S} {440, 967}

\bibitem[\protect\citeauthoryear{{Seifahrt}, {Reiners}, {Almaghrbi}  \&
  {Basri}}{{Seifahrt} et~al.}{2010}]{2010A&A...512A..37S}
{Seifahrt} A.,  {Reiners} A.,  {Almaghrbi} K.~A.~M.,   {Basri} G.,  2010,
  \mn@doi [\aap] {10.1051/0004-6361/200913368}, \href
  {http://cdsads.u-strasbg.fr/abs/2010A%26A...512A..37S} {512, A37}

\bibitem[\protect\citeauthoryear{{Skrutskie} et~al.,}{{Skrutskie}
  et~al.}{2006}]{2006AJ....131.1163S}
{Skrutskie} M.~F.,  et~al., 2006, \mn@doi [\aj] {10.1086/498708}, \href
  {http://cdsads.u-strasbg.fr/abs/2006AJ....131.1163S} {131, 1163}

\bibitem[\protect\citeauthoryear{{Smart}, {Bucciarelli}, {Lattanzi}, {Massone}
  \& {Chiumiento}}{{Smart} et~al.}{1999}]{SMA99A}
{Smart} R.~L.,  {Bucciarelli} B.,  {Lattanzi} M.~G.,  {Massone} G.,
  {Chiumiento} G.,  1999, \aap, 348, 653

\bibitem[\protect\citeauthoryear{{Smart} et~al.,}{{Smart}
  et~al.}{2013}]{2013MNRAS.433.2054S}
{Smart} R.~L.,  et~al., 2013, \mn@doi [\mnras] {10.1093/mnras/stt876}, \href
  {http://adsabs.harvard.edu/abs/2013MNRAS.433.2054S} {433, 2054}

\bibitem[\protect\citeauthoryear{{Smart} et~al.,}{{Smart}
  et~al.}{2017a}]{2017MNRAS.468.3764S}
{Smart} R.~L.,  et~al., 2017a, \mn@doi [\mnras] {10.1093/mnras/stx723}, \href
  {http://cdsads.u-strasbg.fr/abs/2017MNRAS.468.3764S} {468, 3764}

\bibitem[\protect\citeauthoryear{{Smart}, {Marocco}, {Caballero}, {Jones},
  {Barrado}, {Beam{\'{\i}}n}, {Pinfield}  \& {Sarro}}{{Smart}
  et~al.}{2017b}]{2017MNRAS.469..401S}
{Smart} R.~L.,  {Marocco} F.,  {Caballero} J.~A.,  {Jones} H.~R.~A.,  {Barrado}
  D.,  {Beam{\'{\i}}n} J.~C.,  {Pinfield} D.~J.,   {Sarro} L.~M.,  2017b,
  \mn@doi [\mnras] {10.1093/mnras/stx800}, \href
  {http://cdsads.u-strasbg.fr/abs/2017MNRAS.469..401S} {469, 401}

\bibitem[\protect\citeauthoryear{{Spergel} et~al.,}{{Spergel}
  et~al.}{2015}]{2015arXiv150303757S}
{Spergel} D.,  et~al., 2015, preprint, \href
  {http://cdsads.u-strasbg.fr/abs/2015arXiv150303757S} {} (\mn@eprint {arXiv}
  {1503.03757})

\bibitem[\protect\citeauthoryear{{Stephens} \& {Leggett}}{{Stephens} \&
  {Leggett}}{2004}]{2004PASP..116....9S}
{Stephens} D.~C.,  {Leggett} S.~K.,  2004, \mn@doi [\pasp] {10.1086/381135},
  \href {http://cdsads.u-strasbg.fr/abs/2004PASP..116....9S} {116, 9}

\bibitem[\protect\citeauthoryear{{Theissen}}{{Theissen}}{2018}]{2018ApJ...862..173T}
{Theissen} C.~A.,  2018, \mn@doi [\apj] {10.3847/1538-4357/aaccfa}, \href
  {http://cdsads.u-strasbg.fr/abs/2018ApJ...862..173T} {862, 173}

\bibitem[\protect\citeauthoryear{{Thompson} et~al.,}{{Thompson}
  et~al.}{2013}]{2013PASP..125..809T}
{Thompson} M.~A.,  et~al., 2013, \mn@doi [\pasp] {10.1086/671426}, \href
  {http://cdsads.u-strasbg.fr/abs/2013PASP..125..809T} {125, 809}

\bibitem[\protect\citeauthoryear{{Tinney}, {Burgasser}  \&
  {Kirkpatrick}}{{Tinney} et~al.}{2003}]{2003AJ....126..975T}
{Tinney} C.~G.,  {Burgasser} A.~J.,   {Kirkpatrick} J.~D.,  2003, \mn@doi [\aj]
  {10.1086/376481}, \href {http://cdsads.u-strasbg.fr/abs/2003AJ....126..975T}
  {126, 975}

\bibitem[\protect\citeauthoryear{{Tinney}, {Burgasser}, {Kirkpatrick}  \&
  {McElwain}}{{Tinney} et~al.}{2005}]{2005AJ....130.2326T}
{Tinney} C.~G.,  {Burgasser} A.~J.,  {Kirkpatrick} J.~D.,   {McElwain} M.~W.,
  2005, \mn@doi [\aj] {10.1086/491734}, \href
  {http://cdsads.u-strasbg.fr/abs/2005AJ....130.2326T} {130, 2326}

\bibitem[\protect\citeauthoryear{{Vrba} et~al.,}{{Vrba}
  et~al.}{2004}]{2004AJ....127.2948V}
{Vrba} F.~J.,  et~al., 2004, \mn@doi [\aj] {10.1086/383554}, \href
  {http://cdsads.u-strasbg.fr/abs/2004AJ....127.2948V} {127, 2948}

\bibitem[\protect\citeauthoryear{{Wang} et~al.,}{{Wang}
  et~al.}{2018}]{2018PASP..130f4402W}
{Wang} Y.,  et~al., 2018, \mn@doi [\pasp] {10.1088/1538-3873/aaacc5}, \href
  {http://cdsads.u-strasbg.fr/abs/2018PASP..130f4402W} {130, 064402}

\bibitem[\protect\citeauthoryear{{Weinberger}, {Boss}, {Keiser},
  {Anglada-Escud{\'e}}, {Thompson}  \& {Burley}}{{Weinberger}
  et~al.}{2016}]{2016AJ....152...24W}
{Weinberger} A.~J.,  {Boss} A.~P.,  {Keiser} S.~A.,  {Anglada-Escud{\'e}} G.,
  {Thompson} I.~B.,   {Burley} G.,  2016, \mn@doi [\aj]
  {10.3847/0004-6256/152/1/24}, \href
  {http://cdsads.u-strasbg.fr/abs/2016AJ....152...24W} {152, 24}

\bibitem[\protect\citeauthoryear{{West}, {Hawley}, {Bochanski}, {Covey},
  {Reid}, {Dhital}, {Hilton}  \& {Masuda}}{{West}
  et~al.}{2008}]{2008AJ....135..785W}
{West} A.~A.,  {Hawley} S.~L.,  {Bochanski} J.~J.,  {Covey} K.~R.,  {Reid}
  I.~N.,  {Dhital} S.,  {Hilton} E.~J.,   {Masuda} M.,  2008, \mn@doi [\aj]
  {10.1088/0004-6256/135/3/785}, \href
  {http://cdsads.u-strasbg.fr/abs/2008AJ....135..785W} {135, 785}

\bibitem[\protect\citeauthoryear{{Wielen}}{{Wielen}}{1977}]{1977A&A....60..263W}
{Wielen} R.,  1977, \aap, \href
  {http://adsabs.harvard.edu/abs/1977A%26A....60..263W} {60, 263}

\bibitem[\protect\citeauthoryear{{Wilson}, {Kirkpatrick}, {Gizis}, {Skrutskie},
  {Monet}  \& {Houck}}{{Wilson} et~al.}{2001}]{2001AJ....122.1989W}
{Wilson} J.~C.,  {Kirkpatrick} J.~D.,  {Gizis} J.~E.,  {Skrutskie} M.~F.,
  {Monet} D.~G.,   {Houck} J.~R.,  2001, \mn@doi [\aj] {10.1086/323134}, \href
  {http://cdsads.u-strasbg.fr/abs/2001AJ....122.1989W} {122, 1989}

\bibitem[\protect\citeauthoryear{{Wilson}, {Miller}, {Gizis}, {Skrutskie},
  {Houck}, {Kirkpatrick}, {Burgasser}  \& {Monet}}{{Wilson}
  et~al.}{2003}]{2003IAUS..211..197W}
{Wilson} J.~C.,  {Miller} N.~A.,  {Gizis} J.~E.,  {Skrutskie} M.~F.,  {Houck}
  J.~R.,  {Kirkpatrick} J.~D.,  {Burgasser} A.~J.,   {Monet} D.~G.,  2003, in
  {Mart{\'{\i}}n} E.,  ed.,  IAU Symposium Vol. 211, Brown Dwarfs. p.~197

\bibitem[\protect\citeauthoryear{{Zapatero Osorio}, {Mart{\'{\i}}n},
  {B{\'e}jar}, {Bouy}, {Deshpande}  \& {Wainscoat}}{{Zapatero Osorio}
  et~al.}{2007}]{2007ApJ...666.1205Z}
{Zapatero Osorio} M.~R.,  {Mart{\'{\i}}n} E.~L.,  {B{\'e}jar} V.~J.~S.,  {Bouy}
  H.,  {Deshpande} R.,   {Wainscoat} R.~J.,  2007, \mn@doi [\apj]
  {10.1086/520673}, \href {http://cdsads.u-strasbg.fr/abs/2007ApJ...666.1205Z}
  {666, 1205}

\bibitem[\protect\citeauthoryear{{Zhang}, {Homeier}, {Pinfield}, {Lodieu},
  {Jones}, {Allard}  \& {Pavlenko}}{{Zhang} et~al.}{2017}]{2017MNRAS.468..261Z}
{Zhang} Z.~H.,  {Homeier} D.,  {Pinfield} D.~J.,  {Lodieu} N.,  {Jones}
  H.~R.~A.,  {Allard} F.,   {Pavlenko} Y.~V.,  2017, \mn@doi [\mnras]
  {10.1093/mnras/stx350}, \href
  {http://cdsads.u-strasbg.fr/abs/2017MNRAS.468..261Z} {468, 261}

\bibitem[\protect\citeauthoryear{{van Leeuwen}}{{van
  Leeuwen}}{2007}]{2007A&A...474..653V}
{van Leeuwen} F.,  2007, \mn@doi [\aap] {10.1051/0004-6361:20078357}, \href
  {http://cdsads.u-strasbg.fr/abs/2007A%26A...474..653V} {474, 653}

\makeatother
\end{thebibliography}

\bsp	
\label{lastpage}
\end{document}